\documentclass[reprint,aps,pre,twocolumn]{revtex4-1}
\usepackage{amsmath,amssymb,mathrsfs}
\usepackage{amsbsy,amsthm,amsfonts}
\usepackage{graphicx}

\allowdisplaybreaks

\begin{document}
\title{Solution of the nonlinear inverse scattering problem by T-matrix completion. II. Simulations}

\author{Howard W. Levinson}

\affiliation{Department of Mathematics, University of Pennsylvania, Philadelphia, Pennsylvania 19104, USA}
\email{levh@sas.upenn.edu}

\author{Vadim A. Markel}

\thanks{On leave from the Department of Radiology, University of Pennsylvania, Philadelphia, Pennsylvania 19104, USA}
\email{vmarkel@fresnel.fr,vmarkel@mail.med.upenn.edu}
\affiliation{Aix-Marseille Universit\'{e}, CNRS, Centrale Marseille, Institut Fresnel UMR 7249, 13013 Marseille, France}

\date{\today}

\begin{abstract}
This is Part II of the paper series on data-compatible T-matrix completion (DCTMC), which is a method for solving nonlinear inverse problems. Part I of the series contains theory and here we present simulations for inverse scattering of scalar waves. The underlying mathematical model is the scalar wave equation and the object function that is reconstructed is the medium susceptibility. The simulations are relevant to ultrasound tomographic imaging and seismic tomography. It is shown that DCTMC is a viable method for solving strongly nonlinear inverse problems with large data sets. It provides not only the overall shape of the object but the quantitative contrast, which can correspond, for instance, to the variable speed of sound in the imaged medium.
\end{abstract}

\maketitle

\section{Introduction}
\label{sec:BS}

This paper is Part II of the series on solving nonlinear inverse scattering problems (ISPs) by data-compatible T-matrix completion (DCTMC). Part I~\cite{PRE_1} contains theory and here we report initial numerical simulations for the three-dimensional diffraction tomography with scalar waves. This problem arises, in particular, in tomographic ultrasound imaging~\cite{bronstein_02_1,devore_05_1} and in seismology~\cite{liu_12_1,jakobsen_12_1,jakobsen_15_1}. We note that DCTMC can be applied in a very similar manner to the problems of inverse electromagnetic scattering~\cite{carney_04_1,belkebir_05_1,belkebir_06_1,bao_07_1,mudry_12_1} and diffuse optical tomography~\cite{boas_01_1,arridge_09_1}. However, in this paper we probe the medium with scalar propagating waves. By doing so, we avoid, on one hand, the additional complexity related to the vectorial nature of electromagnetic fields and, on the other hand, the severe ill-posedness of the ISP associated with the exponential decay of diffuse waves. Therefore, we can focus on the effects of nonlinearity of the ISP. For the same reason, no noise will be added to the simulated data. 

We have studied convergence and computational performance of several modifications of the DCTMC algorithm. We start with the streamlined iteration cycle with both computational shortcuts as described in~\cite{PRE_1} used. This approach minimizes computational time per one iteration of DCTMC. However, it is not the most efficient algorithm in terms of the number of iterations required for convergence. We then implement several modifications and improvements of DCTMC. In particular, we utilize weighted summation to the diagonal for the ``force-diagonalization operator'' ${\mathcal D}[\cdot]$, which is used to reduce the iteratively-updated interaction matrix $V_k$ to a diagonal matrix $D_k$. This makes the use of Computational Shortcut 2 impossible. However, weighted summation to the diagonal uses explicitly the off-diagonal elements of $V_k$ and is, therefore, more in line with the main idea of DCTMC. Correspondingly, the convergence of DCTMC is significantly improved. With all the improvements, the number of required iterations in a representative test case was reduced from $900$ (for the standard algorithm) to $75$. We also demonstrate how physical constraints and checks for sparsity can be embedded in the DCTMC algorithm. Application of these checks can be viewed as a method of regularizing the nonlinear ISP. 

The remainder of this paper is organized as follows. In Sec.~\ref{sec:scal_wave}, we describe the procedure for discretization of the scalar wave equation. Technical details of the numerical procedures, description of the targets and of the source-detector arrangements used are given in Sec.~\ref{sec:add}. Numerical results for the DCTMC algorithm without any modifications are shown in Sec.~\ref{sec:num}. In Sec.~\ref{sec:impr}, we illustrate several methods to improve the rate of convergence of DCTMC. In Sec.~\ref{sec:GN} we show that DCTMC can provide good results when the Gauss-Newton method fails. Finally, Sec.~\ref{sec:disc} contains a brief discussion of the obtained results.

\section{Discretization of the scalar wave equation}
\label{sec:scal_wave}

Consider a scalar field $u({\bf r})$ (e.g., the pressure wave in ultrasound imaging) and the wave equation
\begin{equation}
\label{wave_eq}
\left[\nabla^2 + k^2 \epsilon({\bf r}) \right]u({\bf r}) = - 4\pi k^2 q({\bf r}) \ , 
\end{equation}
where $q({\bf r})$ is the source function, $\epsilon({\bf r}) = 1$ outside of the bounded domain $\Omega$ (the sample) and the factor $-4\pi k^2$ in the right-hand side has been introduced for convenience. We work in the frequency domain and assume that the wave number $k = \omega/c$ is fixed, where $c$ is the velocity of waves in free space, that is, outside of $\Omega$. 

Since we wish to implement a numerical procedure for reconstructing $\epsilon({\bf r})$ from external measurements, the problem must be suitably discretized and the unknown function must be represented by a finite number of degrees of freedom (voxels). To this end, we follow the general approach of the discrete dipole approximation that was originally developed for Maxwell's equations~\cite{purcell_73_1,draine_94_1}. A related method for the scalar diffusion equation was described by us in~\cite{markel_07_3}.

We start by re-writing \eqref{wave_eq} identically as
\begin{equation}
\label{wave_1}
\left(\nabla^2 + k^2 \right) u({\bf r}) = - 4\pi k^2 \left[ \chi({\bf r}) u({\bf r}) + q({\bf r}) \right] \ , 
\end{equation}
\noindent
where
\begin{equation}
\label{chi_def}
\chi({\bf r}) \equiv \frac{\epsilon({\bf r})-1}{4\pi} \ .
\end{equation}
\noindent
This quantity can be referred to as the susceptibility of the medium. Inverting the differential operator in the left-hand side of \eqref{wave_1}, we obtain Lippmann-Schwinger equation
\begin{equation}
\label{u_LS}
u({\bf r}) = u_{\rm inc}({\bf r}) + \int G_0({\bf r},{\bf r}^\prime)\chi({\bf r}^\prime) u({\bf r}^\prime) d^3 r^\prime \ ,
\end{equation}
\noindent
where 
\begin{equation}
\label{u0_def}
u_{\rm inc}({\bf r}) = \int G_0({\bf r}, {\bf r}^\prime) q({\bf r}^\prime) d^3r^\prime 
\end{equation}
\noindent
is the incident field and
\begin{equation}
\label{G0_def}
G_0({\bf r}, {\bf r}^\prime) = k^2 \frac{\exp\left(ik \vert {\bf r} - {\bf r}^\prime \vert \right)}{\vert {\bf r} - {\bf r}^\prime \vert}
\end{equation}
\noindent
is the free-space Green's function of the wave equation, which satisfies
\begin{equation}
\label{G0_SW}
\left(\nabla^2 + k^2 \right) G_0({\bf r},{\bf r}^\prime) = -4\pi k^2 \delta({\bf r} - {\bf r}^\prime) \ .
\end{equation}
\noindent
Note that the second term in the right-hand side of \eqref{u_LS} is the scattered field $u_{\rm scatt}({\bf r})$, that is,
\begin{equation}
\label{us_def}
u_{\rm scatt}({\bf r}) = \int G_0({\bf r},{\bf r}^\prime)\chi({\bf r}^\prime) u({\bf r}^\prime) d^3 r^\prime \ .
\end{equation}
\noindent
The total field is given by a sum of the incident and scattered components, $u = u_{\rm inc} + u_{\rm scatt}$.

We now proceed with discretization of the integral equation \eqref{u_LS}. Let the sample be rectangular and discretized into cubic voxels ${\mathbb C}_n$ ($n=1,\ldots,N_v$) of volume $h^3$ each, where $N_v$ is the total number of voxels. We then make the following approximation~\cite{fn1}:
\begin{equation}
\label{disc-approx}
\chi({\bf r}) = \chi_n \ \ \ \ {\rm AND} \ \ \ \ u({\bf r}) = u_n \ \ \ \ {\rm IF} \ \ \ \ {\bf r} \in {\mathbb C}_n \ .
\end{equation}
\noindent
Here $\chi_n$ and $u_n$ are constants. Obviously, with this approximation used, \eqref{u_LS} can not hold for all values of ${\bf r}$.  However, by restricting attention to the points ${\bf r} = {\bf r}_n$, where ${\bf r}_n$ are the centers of the respective voxels ${\mathbb C}_n$, we obtain a set of algebraic equations
\begin{equation}
\label{u_LS_1}
u_n = e_n + \sum_{m=1}^{N_v} \chi_m u_m \int_{{\bf r} \in {\mathbb C}_m} G_0({\bf r}_n,{\bf r}) d^3 r \  , 
\end{equation}
\noindent
where
\begin{equation}
\label{e_def}
e_n \equiv u_{\rm inc}({\bf r}_n) \ .
\end{equation}
\noindent
Equation \eqref{u_LS_1} can be reasonably expected to have a solution, which then gives a discrete approximation to the continuous field $u({\bf r})$.

In principle, we can compute the integrals in the right-hand side of \eqref{u_LS_1} analytically or numerically with any degree of precision. However, doing so is not practically useful because there is already an approximation involved in writing \eqref{u_LS_1}. Therefore, the conventional approach is to approximate the integrals by expressions that are easy to compute if $G_0$ is known analytically. To obtain such expressions, we consider the two cases $n \neq m$ and $n = m$ separately. 

For $n \neq m$, the standard approximation is
\begin{equation}
\label{int_n=/=m}
\int_{{\bf r} \in {\mathbb C}_m} G_0({\bf r}_n,{\bf r}) d^3 r \approx h^3 G_0({\bf r}_n,{\bf r}_m) \ , \ \ \ n \neq m \ .
\end{equation}
\noindent
This approximation is not very good for neighboring cells but generally believed to be adequate for cubic lattices (it would be precise for spherical region). However, it should not be applied unaltered for orthorhombic lattices with primitive lattices of non-equal length~\cite{smunev_15_1}. 

For $n=m$, the integrand contains a singularity and a more careful evaluation of the integral is required. It is standard to assume that $kh \ll 1$ (otherwise, the discretization is not a valid approximation to the underlying differential equation) so that the exponent in \eqref{G0_def} can be approximated as 
\begin{equation}
\label{exp_exp}
\exp{\left(ik\vert {\bf r} - {\bf r}^\prime\vert \right)} \approx 1 + ik \vert {\bf r} - {\bf r}^\prime\vert \ . 
\end{equation}
\noindent
We then obtain
\begin{align}
&\int_{{\bf r} \in {\mathbb C}_n} G_0({\bf r}_n,{\bf r}) d^3 r \approx \int_{-h/2}^{h/2} dx \int_{-h/2}^{h/2} dy \int_{-h/2}^{h/2} dz \nonumber \\
&
\times \left(\frac{k^2}{\sqrt{x^2 + y^2 + z^2}} + ik^3 \right) = (kh)^2\left(\xi + ikh \right) \ ,
\label{int_n=m}
\end{align}
\noindent
where 
\begin{equation}
\label{xi_def}
\xi = \log \left( 26 + 15\sqrt{3} \right)-\pi/2 \approx 2.38 \ .
\end{equation}
\noindent
The imaginary part of the right-hand side of \eqref{int_n=m} is the first non-vanishing radiative correction to the voxel polarizability. As is known for the electromagnetic case, accounting for this correction is important to ensure energy conservation of the scattering process~\cite{draine_88_1,markel_92_1}. Note that making higher-order approximations in \eqref{exp_exp} is known to produce higher-order {\em dynamic corrections}. For the electromagnetic case, the relevant results are given in~\cite{lakhtakia_92_2,draine_93_1}. Also, an alternative estimate of the parameter $\xi$ can be obtained if we replace integration over a cubic voxel by integration over a ball of equal volume. This approach results in a much simpler integral and does not affect the radiative correction; however, it gives a slightly different value for $\xi$, namely, $\xi = (9\pi/2)^{1/3} \approx 2.42$. The two approaches to computing $\xi$ can be applicable to different physical situations. For example, integration over a ball may be more appropriate if we wish to describe scattering by a collection of spherical particles rather than voxelization of a medium on a cubic grid. A review of different definitions of the parameter $\xi$ for the electromagnetic case and a summary of several relevant expressions, including \eqref{xi_def}, are given in~\cite{yurkin_13_1}.

Further, it is convenient to introduce the ''moments''
\begin{equation}
\label{d_def}
d_n \equiv h^3 \chi_n u_n \ . 
\end{equation}
\noindent
In terms of $d_n$, and with the account of the integration results \eqref{int_n=/=m} and \eqref{int_n=m}, the system of equations \eqref{u_LS_1} takes the form
\begin{equation}
\label{u_LS_3}
d_n = \alpha_n \left( e_n + \sum_{m=1}^{N_v} {\it \Gamma}_{nm} d_m \right) \ ,  
\end{equation}
\noindent
where
\begin{equation}
\label{alpha_def}
\alpha_n = \frac{h^3 \chi_n}{1 - (kh)^2(\xi + ikh) \chi_n}
\end{equation}
\noindent
is the polarizability of $n$-th voxel and
\begin{equation}
\label{Gamma_def}
{\it \Gamma}_{nm} = \left(1 - \delta_{nm}  \right) G_0({\bf r}_n,{\bf r}_m) 
\end{equation}
\noindent
is the interaction matrix. Note that ${\it \Gamma}_{nm}$ has zeros on the diagonal.

If all polarizabilities $\alpha_n$ are known, then equation \eqref{u_LS_3} is the statement of the forward scattering problem. Namely, given the incident field in all voxels, $e_n$, find the moments $d_n$ by solving \eqref{u_LS_3}; then use the formula 
\begin{equation}
\label{us_disc}
u_{\rm scatt}({\bf r}_d) = \sum_{n=1}^{N_v} G_0({\bf r}_d, {\bf r}_n) d_n
\end{equation}
\noindent
[a discretized version of \eqref{us_def}] to find the scattered field at an arbitrary point of observation ${\bf r}_d$. Obviously, the forward problem is linear with respect to the unknowns, $d_n$. 

The inverse problem is stated differently and is, in general, nonlinear. Our goal is to use multiple measurements of the scattered field $u_{\rm scatt}({\bf r}_d)$ due to multiple external point sources of the form $q({\bf r}) = \delta({\bf r} - {\bf r}_s)$, where ${\bf r}_d$ and ${\bf r}_s$ are the positions of the detector and the source, to find all the voxel polarizabilities $\alpha_n$. In the matrix form, the ISP is stated as follows. Let us define the interaction matrix $V$ as the $N_v \times N_v$ matrix that contains  $\alpha_n$ on the diagonal and zeros elsewhere. Then we can re-write \eqref{u_LS_3} in matrix notations as
\begin{equation}
\label{d_E_matrix}
\vert d \rangle = V \left( \vert e \rangle + {\it \Gamma} \vert d \rangle \right) \ , 
\end{equation}
\noindent
where $\vert d \rangle$ and $\vert e \rangle$ are the vectors of the length $N_v$ with the elements $d_n$ and $e_n$, respectively. The formal solution to \eqref{d_E_matrix} is
\begin{equation}
\label{d_E}
\vert d \rangle = (I - V{\it \Gamma})^{-1} V \vert e \rangle = T[V]\vert e \rangle \ , 
\end{equation}
\noindent
where 
\begin{equation}
\label{T_E}
T[V] \equiv (I - V{\it \Gamma})^{-1} V
\end{equation}
\noindent
is the T-matrix. Let us assume that we measure the scattered field at a set of points ${\bf r}_{dk} \in \Sigma_d$, $k=1,\ldots,N_d$, and let $f_k = u_{\rm scatt}({\bf r}_{dk})$. Then $\vert f \rangle$ is a vector of the length $N_d$ and we can use \eqref{us_disc} to write $\vert f \rangle = A \vert d \rangle$, where $A_{kn} = G_0({\bf r}_{dk}, {\bf r}_n)$.
Similarly, we have $\vert e \rangle = B \vert q \rangle$, where $B_{nl} = G_0({\bf r}_n,{\bf r}_{sl})$ and ${\bf r}_{sl} \in \Sigma_s$, $l=1,\ldots, N_s$ is the set of points from which we illuminate the medium. By turning on one source  at a time and by measuring the scattered field at all points ${\bf r}_{dk}$ for each source, we measure all elements of the data matrix ${\it \Phi}_{kl}$. As follows from \eqref{d_E}, the data matrix satisfies the equation
\begin{equation}
\label{PRE_matrix}
A T[V] B = {\it \Phi} \ .
\end{equation}
\noindent
This is equivalent to Eqs.~4 and 8 of~\cite{PRE_1}. As in \cite{PRE_1}, the matrices $A$ and $B$ are obtained directly by sampling the Green's function $G_0({\bf r},{\bf r}^\prime)$, for which we have given a specific expression \eqref{G0_def} that corresponds to the physical problem considered. The only fine (and slightly nontrivial) point here is the way in which the matrix ${\it \Gamma}$ was obtained. For $n\neq m$, we still have ${\it \Gamma}_{nm} = G_0({\bf r}_n,{\bf r}_m)$, i.e., the off-diagonal matrix elements were obtained by sampling $G_0$. However, the diagonal elements of ${\it \Gamma}$ can not be obtained by straightforward sampling due to the singularity of $G_0$. The discretization procedure described above results in ${\it \Gamma}_{nn} = 0$ provided that we use the renormalized polarizabilities $\alpha_n$ \eqref{alpha_def} to quantify the response of individual voxels to external excitation.

The above approach to obtaining the renormalized polarizabilities is rather standard in physics and goes back to the Lorentz's idea of local-field correction (in electromagnetics). However, it is useful to keep in mind that the local-field correction and the related renormalization are mathematically related to the singularity of the free-space Green's function. From the purely algebraic point of view, we can point out that the renormalization in question is a special case of the identical transformation
\begin{equation*}
(I - V{\it \Gamma})^{-1} V = (I - V^\prime{\it \Gamma}^\prime)^{-1} V^\prime
\end{equation*}
\noindent
where $V^\prime = PV$, ${\it \Gamma}^\prime = {\it \Gamma} - V^{-1} (I - P^{-1})$ and $P$ is any invertible matrix (although $V^{-1}$ appears in the above expression, invertibility of $V$ is not really required). In particular, if ${\it \Gamma}$ has the diagonal part $D$, then we can take $P = (I-VD)^{-1}$ (assuming that this inverse exists) and the renormalized matrix ${\it \Gamma}^\prime$ will have a zero diagonal. It also should be pointed out that the renormalized polarizabilities derived here are adequate for cubic lattices but require modification in the case of orthorhombic lattices with primitive vectors of unequal length~\cite{smunev_15_1}. 

\section{Details of the numerical algorithms used}
\label{sec:add}

\subsection{Definition of the unknowns}
\label{sec:add.cfu}

The above inverse problem has been formulated in terms of the unknown voxel polarizabilities $\alpha_n$. However, the physical quantity of interest is the voxel permittivity $\epsilon_n$ or the susceptibility $\chi_n$. To relate the latter to the former, we can use \eqref{chi_def} and \eqref{alpha_def} to obtain in a straightforward manner
\begin{equation}
\label{chi_alpha}
\chi_n = \frac{\alpha_n/h^3}{1 + (kh)^2(\xi + i k h) (\alpha_n/h^3)} \ .
\end{equation}
\noindent
The permittivities $\epsilon_n$ can be trivially related to $\chi_n$ by using \eqref{chi_def}, and we will focus on the latter quantities in the remainder of this paper.

Equation~\eqref{chi_alpha} is an analog of the Maxwell Garnett formula written for scalar waves (with the account of the first non-vanishing radiative correction) and its inverse is the Clausius-Mossotti relation. It can be seen that the relation between $\chi_n$ and $\alpha_n$ is itself nonlinear. In other words, we have removed {\em some} nonlinearity of the ISP analytically. The nonlinearity in question is solely due to the self-interaction of a given voxel and it was accounted for by the procedure of renormalization of the polarizability. In other words, we can view the numerator $h^3\chi_n$ in the right-hand side of \eqref{alpha_def} as the bare polarizability and the complete expression as the renormalized polarizability. However, the nonlinearity of the ISP that is due to the interaction of different voxels can not be removed so simply and we will have to deal with it numerically.

Still, the above discussion gives some validity (although not a rigorous proof) to the proposition that it is advantageous to formulate the ISP in terms of the polarizabilities $\alpha_n$ rather than in terms of the susceptibilities $\chi_n$. Indeed, consider the case when there is only one voxel and we wish to determine its properties (either $\alpha$ or $\chi$) from external measurements by means of some generic iterative scheme that is applicable to a more general problem (e.g., involving many interacting voxels). If we view $\alpha$ as the fundamental unknown, we arrive at a well-posed linear equation of the form $A\alpha = b$ ($A\neq 0$ and $b$ are numbers), which can be solved in just one iteration. However, if we view $\chi$ as the fundamental unknown, then we will be solving iteratively an equation of the form $A\chi/(1 - \beta \chi) = b$, which can take several iterations depending on the value of the coefficient $\beta$. 

In the simulations reported below, we formulate the model and display the reconstructions in terms of the susceptibilities $\chi_n$, which we assume to be real-valued. However, to generate the data function, we compute the set of $\alpha_n$'s according to \eqref{alpha_def}. Then we ''pretend'' that $\alpha_n$'s are unknown and, viewing these quantities as the fundamental unknowns, solve the ISP. We then convert the reconstructed values of $\alpha_n$'s to $\chi_n$'s by using \eqref{chi_alpha} and display the latter quantities in all figures.

\subsection{Application of physical constraints}
\label{sec:add.pcs}

As discussed in \cite{PRE_1}, the iterative procedure of DCTMC can benefit from applying physical constraints to the unknowns as a form of regularization. The constraints are usually based on general physical principles but can also accommodate some {\em a priori} knowledge about the medium. In this work, only fundamental constraints are used. Specifically, if the medium is passive (that is, non-amplifying), then we know that ${\rm Im}\chi_n \geq 0$. If it is also known that the medium is transparent (non-absorbing), then ${\rm Im}\chi_n = 0$. If we view the polarizabilities $\alpha_n$ as the fundamental unknowns in the iterative DCTMC process, we can apply this physical constraint in the following manner. We first notice that
\begin{equation}
\label{Im_1_over_alpha}
{\rm Im}\left(\frac{h^3}{\alpha_n}\right) = -\left[ \frac{{\rm Im}\chi_n}{\vert \chi_n \vert^2} + (kh)^3 \right] \leq -(kh)^3 \ .
\end{equation}
\noindent
Let's say, a numerical iteration of DCTMC has produced a set of $\alpha_n$. To enforce the condition \eqref{Im_1_over_alpha}, we can apply the following transformation:
\begin{equation}
\label{phys_constraint_gen}
\alpha_n \longrightarrow \frac{1}{{\rm Re}(1/\alpha_n) - i \max \left[ -{\rm Im}(1/\alpha_n) , k^3 \right] } \ .
\end{equation}
\noindent
If, in addition, we know that the sample is non-absorbing so that ${\rm Im}\chi_n=0$ for all voxels and strict equality in the last relation in \eqref{Im_1_over_alpha} holds, then we can apply an even sharper constraint by writing 
\begin{equation}
\label{phys_constraint_nonabs}
\alpha_n \longrightarrow \frac{1}{{\rm Re}(1/\alpha_n) - i k^3 } \ .
\end{equation}
\noindent
In the simulations reported below, we have assumed that the sample is transparent and used the constraint \eqref{phys_constraint_nonabs} at each iteration of DCTMC. We note that in many cases considered, DCTMC produces very similar results even without applying the constraint, but then the convergence is somewhat slower. The only instances in which we found that the physical constraint is critical were the cases with very strong nonlinearity. 

\subsection{Account of sparsity}
\label{sec:add.spr}

In many practical cases we can expect the target to be in some sense sparse. This means that many of the voxels have zero (or relatively small) susceptibilities $\chi_n$, but we do not know {\em a priori} where these voxels are located or how many such voxels exist in the computational domain. We will refer to such voxels as ''noninteracting'' as the moments $d_n$ of these voxels are relatively small, do not interact effectively with the moments of other voxels, and produce a negligible input to the scattered field.

In DCTMC, it is possible to take the sparsity into account in an adaptive manner without actually knowing whether the target is sparse or not or where the non-interacting voxels are located. In the simulations of Sec.~\ref{sec:num} below, we have used the following rather {\em ad hoc} algorithm:

\begin{enumerate}

\item[1:] Run $50$ iterations normally.

\item[2:] Then every $20$ iterations check whether some susceptibilities $\chi_n$ satisfy $\vert \chi_n \vert < \chi_{\rm max}/100$, where $\chi_{\rm max} = \max_n \vert \chi_n \vert$. 

\item[3:] If a given voxel satisfies the above condition $3$ checks in a row, the corresponding $\chi_n$ is set to zero.

\item[4:] The voxels with zero $\chi_n$ (as determined in the previous step) are declared to be non-interacting and are excluded from the computational domain. When this happens, we repeat the initial DCTMC setup, but now for a smaller number of interacting voxels $N_v$. This results in a smaller computational time per a subsequent iteration.

\item[5:] The process is repeated with the following modifications. After $200$ iterations, checks are made every $10$ iterations, and after $400$ iterations, the relative threshold for determining a non-interacting voxel is reduced to the factor of $60$, and after $600$ iterations the relative threshold is reduced to the factor of $40$.

\end{enumerate}

Note that all integer constants used in the above algorithm are adjustable. In Sec.~\ref{sec:impr}, we have used somewhat different values of these constants (as noted in each particular case) and in some simulations we did not use sparsity checks at all.

The procedure of selecting the noninteracting voxels can be described as iterative ''roughening'' of the target. Any iterative numerical reconstruction is expected to produce small but nonzero values in the regions of the computational domain where the reconstructed function is really zero. Keeping these small values in subsequent iterations is nothing but a computational burden. The procedure described here removes this burden without affecting the final images in a significant way as long as the noninteracting voxels are not assigned incorrectly. In the simulations shown below, such incorrect assignment occurred in the strong nonlinearity regime. We note that such occurrences can be easily ''diagnosed'' by monitoring the error of the matrix equation \eqref{PRE_matrix} as is explained below. If an incorrect assignment does occur, one can potentially alleviate the problem by adjusting the constants used by the algorithm or by not using sparsity checks at all.

Finally note that the homogeneous background against which the roughening is performed can be different from zero. Thus, in optical tomography, it is conventional to assume that the background properties of the medium are known yet different from those of free space~\cite{arridge_09_1}. In this case, a more complicated Green's function $G_0$ must be used. We emphasize that $G_0$ can be computed analytically for many regular geometries of the background medium. In this paper, we have taken the homogeneous background to be free space (vacuum) with the sole purpose of being able to use the mathematically-simple function given in \eqref{G0_def}.

\subsection{Model targets and arrangement of sources and detectors}
\label{sec:add.mod}

We have used three kinds of targets in the simulations, which we refer to as large, small and tiny. A target of a particular kind has always the same ''shape'' but can have varying degrees of contrast. The contrast is defined in terms of the susceptibility $\chi({\bf r})$. Mathematically, this means that $\chi({\bf r})$ for a given target can be written as 
\begin{equation}
\label{shape}
\chi({\bf r}) = \chi_0 \Theta({\bf r}) \ , 
\end{equation}
\noindent
where $0 \leq \Theta({\bf r}) \leq 1$ is the shape function (always the same for a target of a given kind) and $\chi_0 > 0$ is the variable amplitude. Obviously, the larger is $\chi_0$, the stronger is the nonlinearity of the ISP. Note that all reconstructions shown below display the ratio $\chi_n /\chi_0$, which, ideally, should coincide with the shape function. However, we did not use any {\em a priori} knowledge about $\chi_0$ to obtain the reconstructions. The normalization to $\chi_0$ was performed {\em a posteriori}, which allowed us to use the same color scale in all figures.

The majority of simulations shown below are performed for the small target, which is discretized on a $16 \times 16 \times 9$ grid ($N_v = 2,304$). DCTMC reconstructions were also demonstrated for the large target, which is discretized on a $30 \times 30 \times 15$ grid ($N_v = 13,500$). The tiny target ($8 \times 8 \times 4$, $N_v=256$) is used only in Sec.~\ref{sec:GN} below for the purpose of comparison of DCTMC with the Gauss-Newton methods since the latter did not yield useful results for the small target and were too computationally inefficient to be applied for the large target.

\begin{figure}
% FIG 1	
\centerline{\includegraphics[width=8.2cm]{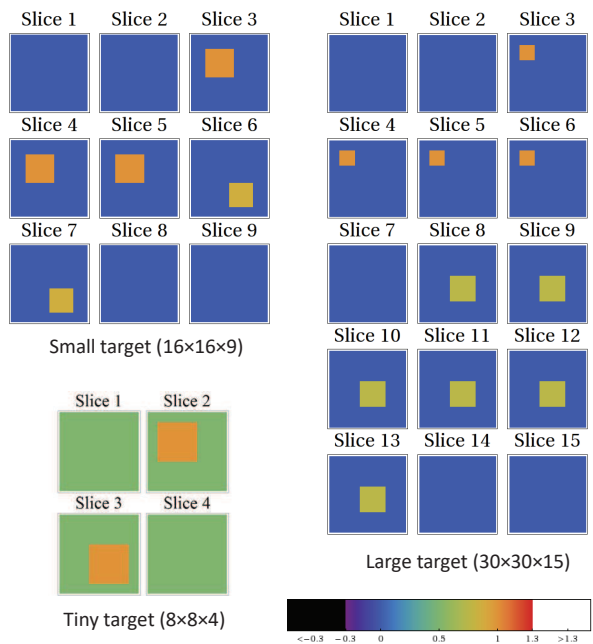}}
\caption{\label{fig:Targets} The shapes of all three targets and the color scheme used in all figures below. Even though all model shape functions satisfy (by definition) the condition $0 \leq \Theta \leq 1$, the color scale can be used to represent any quantity in the range $[-0.3,1.3]$. The allowances are made because the reconstructed values of $\chi_n/\chi_0$ can become negative or larger than unity. Reconstructed values that are smaller than $-0.3$ will be shown by the uniform black color and values that are larger than $1.3$ will be shown as white.}
\end{figure}

The shapes of all targets are shown in Fig.~\ref{fig:Targets}. It can be seen that the small target contains two rectangular inclusions in a homogeneous zero background. One inclusion is of the size $6 \times 6\times 3$ (in units of $h$) and has $\Theta = 1.0$ and the second inclusion is of the size $5 \times 5 \times 2$ and has $\Theta = 0.857$. Note that the smaller and the larger inclusions touch at one corner. The large target contains one rectangular inclusion of the size $6 \times 6 \times 4$ and with $\Theta = 1$ and another inclusion of the size $10 \times 10 \times 6$ and with $\Theta = 0.75$. These two inclusions do not quite touch. The tiny target contains an inhomogeneity in the form of two squares of the size $4 \times 4 \times 1$ in located in the two interstitial slices, as is shown in the figure. The two square regions overlap in the central $2\times 2$ area when viewed from top. In terms of the shape function, we have $\Theta({\bf r})=1$ inside the inhomogeneity and $\Theta({\bf r})=0.5$ in the background. Thus, all voxels in the tiny target are different from free space. 

It can be remarked that the tiny target does not represent an accurate discretization of the underlying differential equation. However, it poses a nonlinear ISP in its own right, and various inversion methods can be tested using this model.

The relation between the discretization step $h$ and the free-space wave number $k$ is $kh=0.2$ in all cases. We can use this numerical value to estimate roughly the degree of nonlinearity of the ISP. We generally expect the ISP to be very nonlinear when the phase shift between two waves -- one propagating through an inhomogeneity (say, in the $z$-direction) and another propagating through the background medium -- becomes of the order of $\pi/2$ or larger. This phase shift is given by
\begin{equation}
\Delta \varphi = (k h) n \left( \sqrt{1 + 4\pi\chi_0 \Theta_i} - 1 \right) \ ,
\end{equation}
\noindent
where $n$ is the inhomogeneity depth in units of $h$, $\chi_0$ is the contrast, and $\Theta_i$ is the value of the shape function inside the inhomogeneity. For example, the largest contrast we used for the small target is $\chi_0 = 1.75$. The shape function inside the inhomogeneity that is $n=3$ voxels deep is $\Theta_i = 0.857$. This corresponds to $\Delta \varphi \approx 0.66\pi$. The nonlinearity in this case is expected to be very strong. Of course, this analysis does not take into account multiple scattering between different inhomogeneities. In fact, we'll see that the linearized reconstructions break down for a much smaller contrast, i.e., at $\chi_0 \approx 0.1$.

We now turn to the source-detector arrangements. 

In the case of the small target, the mesh of sources is a $22 \times 22$ rectangular grid with the same spacing $h$ as was used to discretize the sample. The plane of sources is centered symmetrically near one of the $16\times 16$ faces of the sample so that there are three rows and columns of sources extending past the sample surface in each direction. The plane of sources is displaced by $h/2$ from the physical surface of the sample. The mesh of detectors is identical in dimension and located on the other side of the sample.  Regarding the size of the data set, we have $N_s=N_d=484$ and the total number of data points is $N_s N_d = 234,256$. Note that the additional rows and columns of sources and detectors extending past the sample surface are needed to achieve the best possible linear reconstruction in the weak nonlinearity regime. Adding even more rows and columns into the source/detector meshes does not improve this result any further.

In the case of the large target, the grids of sources and detectors are of the size $38 \times 38$ so that four rows and columns of sources or detectors are extending past the sample surface in each direction. The source/detector meshes are displaced from each of the two $30 \times 30$ faces of the large target by $h/2$, as is also the case for the small target. The data set size is defined by the following numbers: $N_s = N_d = 1,444$ and $N_s N_d = 2,085,136$.

For the tiny target, the grids of sources and detectors were of the size  $10 \times 10$, $(N_d=N_s=100)$ and centered about the sample on all sides. 

\subsection{Methods for linearized reconstruction}
\label{sec:add.lin}

Applying a nonlinear solver to an ISP is meaningful only if the linearized solution to the same problem has failed. Therefore, we compare the results of nonlinear reconstruction to those obtained by linearized inversion. We have used two different approaches to this problem.

First, we have used the traditional approach of computing the pseudoinverse of the matrix $K$ (see Remarks 2 and 3 in~\cite{PRE_1}). This approach leads to a linear problem whose size grows very rapidly with the size of the data set and the number of voxels. While still feasible in the case of the small target, application of this method for the large target is already problematic. Indeed, the matrix $K$ for the large target has the total number of elements $N_s N_d N_v = 28,149,336,000$. Storing a matrix of this size in computer memory requires at least $112{\rm Gb}$ of RAM (in single precision). One can follow the approach of~\cite{ban_13_1} where we have computed the product $K^*K$ iteratively by storing only sufficiently small blocks of $K$ in computer memory. Then the product $K^*K$ was Tikhonov-regularized by applying the operation $K^*K \rightarrow K^*K + \lambda^2 I$ and the resultant system of equations was solved by the conjugate gradient descent. The bottleneck of this approach is the computation of $K^*K$, which can still take very considerable time.

On the other hand, solving the same linearized problem by the method described in Appendix B of~\cite{PRE_1} is a much simpler task, and this approach yields, essentially, the same result. Indeed, in all cases we have considered, the two results were visually indistinguishable. This fact illustrates the proposition that considering the matrices $A$ and $B$ separately rather than combining them into one large matrix $K$ is computationally advantageous even in the linear regime. One can understand this improvement as a result of {\em data reduction}, that is, defining a transform of the data that is smaller in size but still contains all essential information. 
 
All linearized reconstructions shown below were obtained by regularizing and solving the equation $W \vert \upsilon \rangle = \vert \upsilon_{\rm exp} \rangle$ (sec.~7 of \cite{PRE_1}). The positive-definite matrix $W$ of this equation was Tikhonov-regularized by the operation $W \rightarrow W + \lambda^2 I$. The regularization parameter $\lambda$ was adjusted manually to obtain the best linear reconstruction in each case considered. The equation was then solved by the conjugate-gradient descent method. We note that the computational complexity of computing $W$ (which is exactly of the same size as $K^*K$, that is, $N_v \times N_v$) is much smaller than the numerical complexity of computing $K^*K$. Indeed, computing $W$ requires only the singular vectors of the much smaller matrices $A$ and $B$. Therefore, the main computational bottleneck of the linearized inversion is removed in this approach.

We have applied different linearization methods, including first Born, first Rytov and mean-field approximations as described in Appendix A of~\cite{PRE_1}. The choice of a linearization method only affects the way in which the data are computed and not the matrices $K$ or $W$. Extensive simulations have revealed that reconstructions based on first Rytov or mean-field approximations do not provide any noticeable advantages when compared to first Born. First Born and first Rytov reconstructions are compared in Fig.~\ref{fig:small_near} below for illustrative purposes, but in all other cases only first Born-based reconstructions are shown.

\subsection{Error measures}
\label{sec:add.err}

To quantify the convergence of the method, we use normalized root mean square errors $\eta_\chi$ (error of the solution) and $\eta_{\it \Phi}$ (error of the equation). These quantities are defined as
\begin{subequations}
\label{eta_def}
\begin{align}
\label{eta_chi}
& \eta^2_\chi       = \frac{1}{N_v \chi_0^2}     \sum_{n=1}^{N_v} \left[\chi_n^{\rm (Reconstructed)} - \chi_n^{\rm (True)} \right]^2 \ , \\
\label{eta_Phi}
& \eta^2_{\it \Phi} = \frac{1}{N_d N_s \chi_0^2} \sum_{i=1}^{N_d} \sum_{j=1}^{N_s} \left[ {\it \Phi}_{ij} - (ATB)_{ij} \right]^2  \ .
\end{align}
\end{subequations}
\noindent 
Note that $\eta_\chi$ is computed after Step 1 of the streamlined iteration cycle with the use of Computational Shortcut 2 or after Step 2 for the algorithm without the use of Computational Shortcut 2 (as defined in~\cite{PRE_1}). The error $\eta_{\it \Phi}$ is computed after Step 4 of this algorithm. After Step 5, the T-matrix is fully data-compatible and the error in question is zero up to the numerical precision of the computer. It should be kept in mind that $\eta_\chi$ can be computed only if the target is known {\em a priori}, which in practical applications is almost never the case. However, $\eta_{\it \Phi}$ can be computed even if the target is not known {\em a priori}.

\section{Numerical results}
\label{sec:num}

\subsection{Small target}
\label{sec:num.small}

\begin{figure*}
% FIG 2
\centerline{\includegraphics[width=16.4cm]{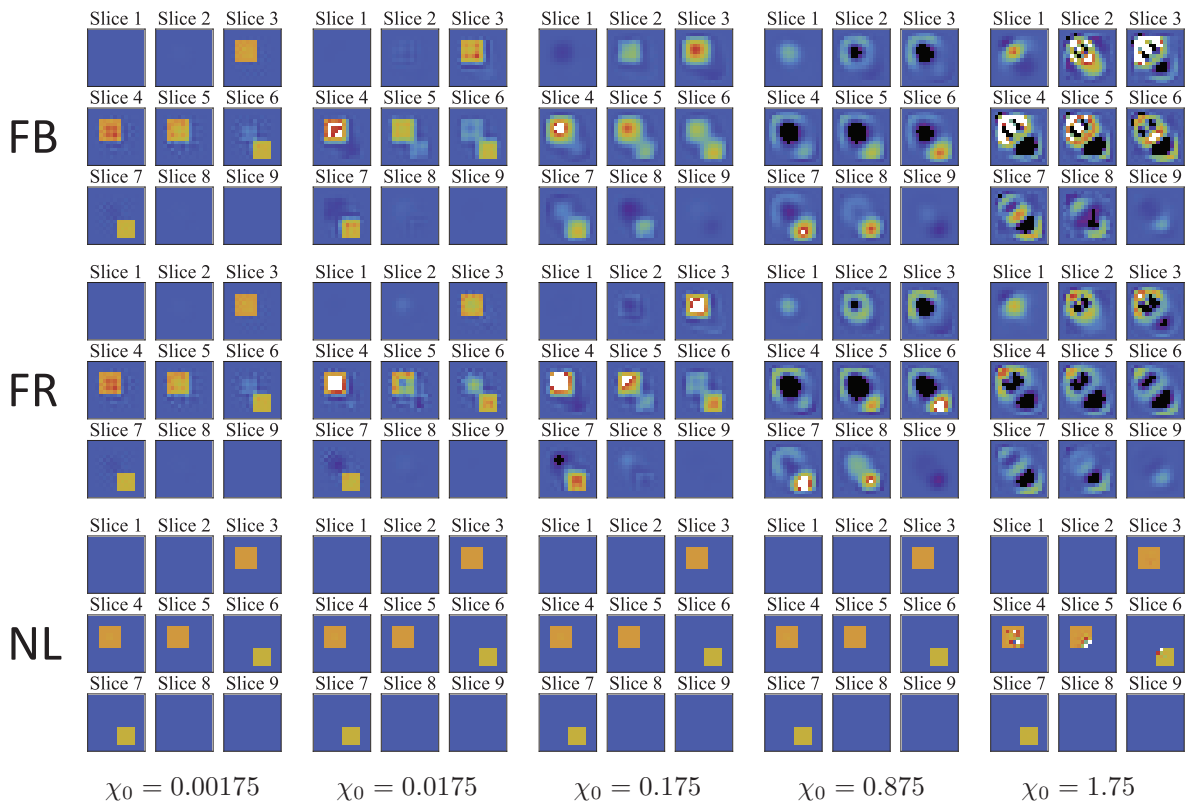}}
\caption{\label{fig:small_near} Linear (top and middle rows, marked FB and FR) and nonlinear (bottom row, marked NL) reconstructions of the small target for different levels of contrast $\chi_0$. The quantity shown by the color scale in each plot is $\chi_n / \chi_0$, where $\chi_n$ is the reconstructed susceptibility of the $n$-th voxel (real under the assumptions used) and $\chi_0$ is the amplitude of the shape function. For the linearized reconstructions, FB denotes first Born approximation and FR denotes first Rytov approximation.}
\end{figure*}

Reconstructed images of the small target after $900$ DCTMC iterations are shown in Fig.~\ref{fig:small_near} for various degrees of contrast. Images obtained by linearized inversion are also shown for comparison. It can be seen that the linearized image reconstruction methods break down between $\chi_0=0.01$ and $\chi_0 = 0.1$. First Rytov approximation does not provide a visible advantage over first Born, and the same is true for the mean-field approximation (data not shown). However, DCTMC yields reconstructions that are close to the correct result for all values of the contrast used. At $\chi_0=1.75$, the DCTMC reconstruction starts to visibly break down. The reason for this is incorrect assignment of three noninteracting voxels, as can be seen from the figure with sufficient magnification. The vicinity of these three voxels is also reconstructed incorrectly. So the breakdown in this case is due to the {\em ad hoc} algorithm for assigning the noninteracting voxels. This problem is not fatal for DCTMC as the algorithm can be adjusted. Note that it is possible to learn whether some noninteracting voxels have been assigned incorrectly by looking at the convergence curve for the error $\eta_{\it \Phi}$ (see Fig.~\ref{fig:small_conv} below). Doing so does not require any {\em a priori} knowledge of the target.

The errors $\eta_\chi$ and $\eta_{\it \Phi}$ for the small target are shown in Fig.~\ref{fig:small_conv} as functions of the iteration number, $i$. We start the discussion of convergence with the error of reconstruction, $\eta_\chi$. The curves $\eta_{\chi}(i)$ all look very similar and are almost independent of $\chi_0$, except for $\chi_0=1.75$. The latter case will be discussed separately. For $\chi_0 < 1.75$, the function $\eta_\chi(i)$ has three distinct convergence regimes, as described below. 

First, there is a region of slow convergence. While DCTMC is in this regime, the reconstructed values $\chi_n$ (or $\alpha_n$) are all very small. Therefore, it can be concluded that the initial iterations simply solve the linearized inverse problem by the iterative algorithm described in~\cite{PRE_1} (Richardson first-order iteration). Naturally, convergence of this iterative process is expected to be slow. In principle, this initial slow convergence regime can be completely avoided by solving the linearized problem directly or by a fast iterative method such as conjugate-gradient descent, and then using the result as the initial guess for DCTMC. As is explained in Sec.~\ref{sec:add.lin}, linearized inversion can be obtained relatively fast with the use of data reduction that is inherent in treating the matrices $A$ and $B$ separately rather than combining them into one large matrix $K$. The computational advantage obtained by this approach is illustrated in Sec.~\ref{sec:impr.lin} below.

Second, there is a region of fast convergence. Assignment of noninteracting voxels occurs in this range of iteration indexes. If the curves are viewed with sufficient magnification, it can be seen that they are not smooth but contain "kinks", that is, points where the slope changes abruptly. This change of slope occurs when one or more noninteracting voxels are determined correctly.

Third, there is the last region in which the rate of convergence is lower than in the second region but still much larger than in the first, slow convergence region. Clearly, the error $\eta_\chi(i)$ decreases in the third region according to an exponential law. In this convergence regime, all noninteracting voxels have been determined correctly and the algorithm improves its estimate of the amplitudes of the remaining interacting voxels. The error $\eta_\chi$ continues to decrease to some very small values. This means that the reconstructions can be made very precise. However, the exponential convergence can not continue indefinitely because the error $\eta_\chi$ can not decrease to an arbitrarily small value; it is bounded from below either by the ill-posedness of the inverse problem or by round-off errors.

The above discussion is valid for the contrasts $\chi_0 < 1.75$. For $\chi_0=1.75$, the pattern is quite different. The reason is that three noninteracting voxels have been  assigned incorrectly in this case. As a result, in the first, slow convergence region, the function $\eta_\chi(i)$ actually increases. These jumps take place when noninteracting voxels are assigned incorrectly (apparently, too early). Then there still exists the fast convergence region, wherein many noninteracting voxels are assigned correctly. Finally, the third, exponential convergence region does not exist for $\chi_0=1.75$. This is so because the incorrectly assigned noninteracting voxels set a relatively large lower bound for the error $\eta_\chi$. 

We next discuss the error $\eta_{\it \Phi}$. This error can be computed even if the target is not known {\em a priori}. First, we note that the dependence $\eta_{\it \Phi}(i)$ can also be classified into three different regimes (slow, fast and exponential), just as it was done for $\eta_\chi(i)$. However, unlike in the case of $\eta_\chi(i)$, the curves $\eta_{\it \Phi}(i)$ depend noticeably on the contrast, $\chi_0$. The dependence is, of course, weak for small $\chi_0$. Thus, the two curves for $\chi_0 = 0.00175$ and $\chi_0=0.0175$ are barely distinguishable. However, the curves for $\chi_0 = 0.0175$, $\chi_0 = 0.175$ and $\chi_0 = 0.875$ can be easily distinguished. The difference is most pronounced in the exponential convergence region. The exponent appears to be the same but the overall factor depends on $\chi_0$. This dependence of $\eta_{\it \Phi}$ on $\chi_0$ is a manifestation of the nonlinearity of the inverse problem. Indeed, it can be easily shown that, in the linear regime, $\chi_0 \rightarrow 0$, $\eta_{\it \Phi}$ is independent of $\chi_0$. We note that the dependence on $\chi_0$ can also be visible in the slow convergence region of the iteration indexes. This does not contradict the previously made observation that, in the slow convergence regime, DCTMC solves the linearized problem iteratively. The reason is that the definition of $\eta_{\it \Phi}$ involves the data matrix ${\it \Phi}$, which, in general, is not proportional to $\chi_0$.

It was mentioned above that the overall factor of the function $\eta_{\it \Phi}(i)$ in the exponential convergence region depends on $\chi_0$. Initially, this factor increases with $\chi_0$. More generally, the curve $\eta_{\it \Phi}(i)$ for $\chi_0=0.875$ goes higher than the curve for $\chi_0=0.175$ and the curve for $\chi_0=0.175$ goes higher than the curve for $\chi_0=0.0175$, etc. However, at larger values of $\chi_0$, this tendency is reversed. Thus, the curve for $\chi_0=1.75$ starts lower than the curve for $\chi_0=0.00175$ even before any noninteracting voxels have been assigned. This non-monotonous dependence of $\eta_{\it \Phi}$ on $\chi_0$ is a manifestation of the rather complicated nonlinearity of the inverse problem and is suggestive of a resonance phenomenon. That is, when we increase the value of $\chi_0$, it passes close to a pole in the complex plane where the T-matrix has a singularity. We note that this non-monotonous dependence was observed for the large target as well (see below).

The curves $\eta_{\it \Phi}(i)$ are rather interesting in the strong nonlinearity case $\chi_0=1.75$. The error $\chi_{\it \Phi}$ initially increases due to the incorrect assignment of the three noninteracting voxels. Then the error drops rapidly when many noninteracting voxels are assigned correctly. Then the error undergoes exponential decay with the same exponent as for the smaller contrasts. Interestingly, the error $\eta_\chi$ in this range of iteration indexes is nearly constant while the error $\eta_{\it \Phi}$ steadily decreases. This is so because $\eta_\chi$ is dominated at this point by the three incorrectly assigned noninteracting voxels while the algorithm still improves the accuracy of $\chi_n$ in the remaining majority of voxels. Finally, the exponential decay of $\eta_{\it \Phi}(i)$ crosses over to exponential growth. Why this happens is not entirely clear; one could expect $\eta_{\it \Phi}(i)$ to plateau. We can conjecture that this crossover to exponential growth occurs due to a complex interplay of the physical constraint (that is still being applied at each iteration) and the use of an incomplete computational domain due to the incorrect exclusion of some of the voxels. We note in passing that the best reconstruction result for $\chi_0=1.75$ would have been obtained if we stopped the iterations at $i\approx 600$.

We conclude that the characteristic behavior of $\eta_{\it \Phi}(i)$ gives one an unambiguous indication that noninteracting voxels have been assigned incorrectly or (in more severe cases) that the obtained DCTMC reconstruction is not useful. 

\begin{figure}
% FIG 3
\centerline{\includegraphics[width=8.2cm]{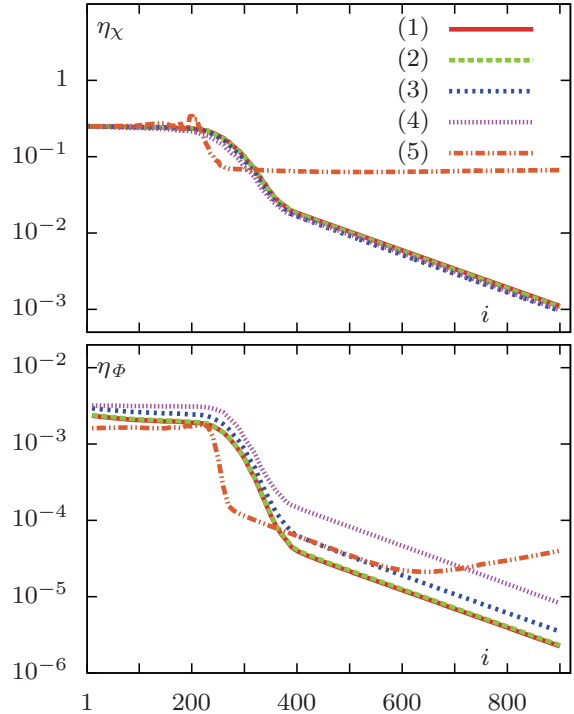}}
\caption{\label{fig:small_conv} Convergence data for the small target. Errors $\eta_\chi$ and $\eta_{\it \Phi}$ are plotted vs the iteration number $i$. Different line types correspond to different contrast $\chi_0$ as follows: $\chi_0=0.00175$ (1), $\chi_0=0.0175$ (2), $\chi_0=0.175$ (3), $\chi_0=0.875$ (4), and $\chi_0=1.75$ (5).}
\end{figure}

\subsection{Large target}
\label{sec:num.large}

\begin{figure*}
% FIG 4
\centerline{\includegraphics[width=16.4cm]{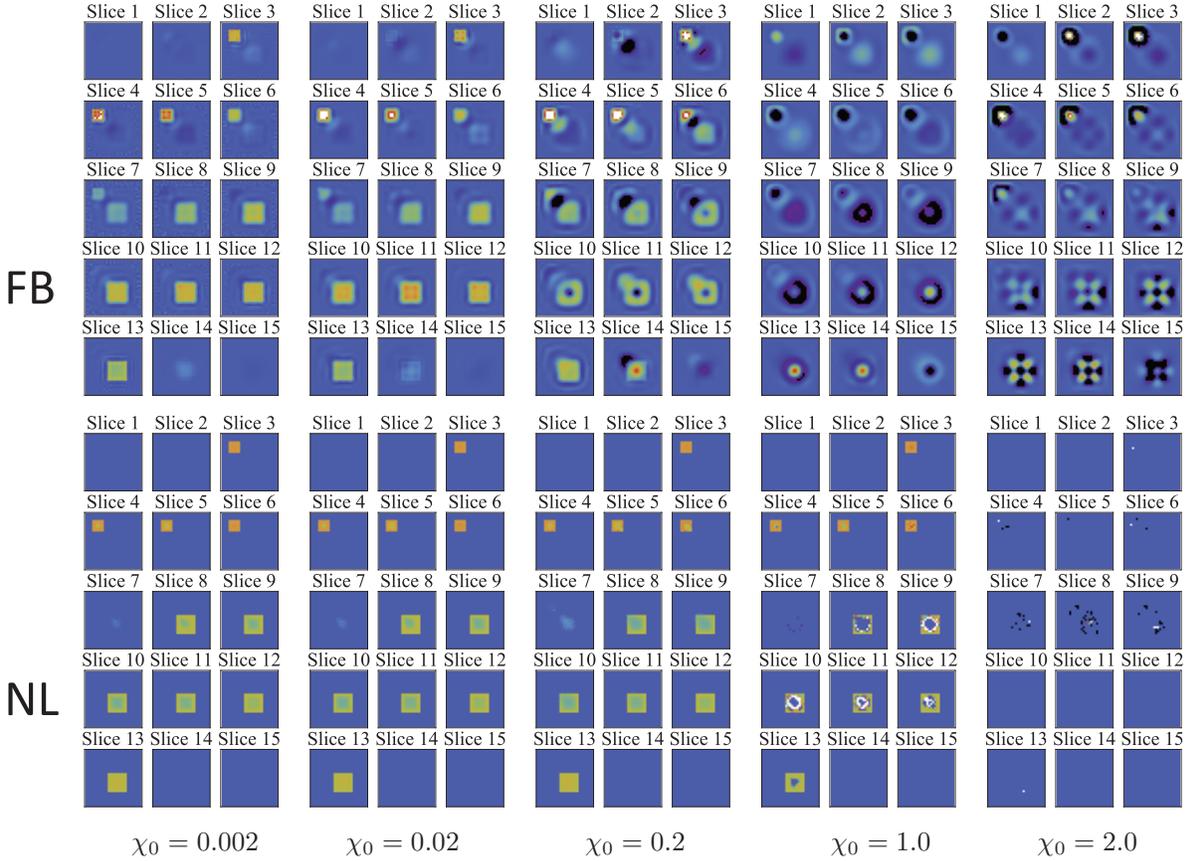}}
\caption{\label{fig:large_near} Same as in Fig.~\ref{fig:small_near}, but for the large target and a somewhat different set of contrasts $\chi_0$. Utilization of first Rytov approximation for linearized reconstruction does not provide any improvements over first Born approximation, and the corresponding results are not shown in this figure.}
\end{figure*}

We now turn to the large target. In Fig.~\ref{fig:large_near}, we show the reconstructions for five different values of the contrast, from $\chi_0 = 0.002$ to $\chi_0=2$. Of course, the computational domain and the size of the inhomogeneities are in this case larger than in the small target and we can expect the onset of nonlinearity to occur at smaller values of the contrast. Also, the linearized inverse problem is more ill-posed because there are voxels in the interior of the large target that are quite far from any source or detector and not effectively probed by incident evanescent waves. Indeed, the linearized reconstructions of the large target are not as good as the similar reconstructions of the small target. We emphasize that these are the best linearized reconstructions we were able to obtain by tuning the Tikhonov regularization parameter. Also, and as was the case for the small target, neither first Rytov nor the mean-field approximation provide a noticeable improvement of the linearized reconstructions (data not shown).

We note however that the DCTMC reconstructions of the large target in the weak nonlinearity regime (e.g., $\chi_0 = 0.002$) are considerably better than the linearized reconstructions. Moreover, it is not possible to improve the image quality of the linearized reconstruction by setting to zero the amplitudes of all voxels that are less in magnitude than, say, $1/40$ of the maximum (recall that $40$ is the smallest thresholding factor used by us in this paper to determine noninteracting voxels in DCTMC). In fact, a voxel with the reconstructed susceptibility $\chi_n = \chi_0/40$ is not visually distinguishable from zero in the color scheme used in this paper. Therefore, the difference in quality between DCTMC and linearized methods is not a trivial consequence of image ''roughening''. The result may appear counter-intuitive. Indeed, in the limit $\chi_0 \rightarrow 0$, the linearized inversion methods and DCTMC are solving exactly the same problem. However, the two approaches involve different regularization methods. In linearized inversions, Tikhonov regularization is used. In DCTMC, regularization is afforded by applying the physical constraint at each iteration. Apparently, the latter approach is much better at reproducing sharp edges.

As we move to the strong nonlinearity regime, the linearized reconstructions break down. At $\chi_0=0.2$, the linearized reconstruction is not useful while DCTMC still provides a quantitatively accurate result. The contrast level $\chi_0=1$ is borderline for DCTMC. It can be seen that the smaller, higher contrast inhomogeneity is still reconstructed correctly. However, the interior region of the larger inhomogeneity is not properly reconstructed (for the most part, underestimated). The boundaries of the larger inhomogeneity are nevertheless clearly visible. Note that we have encountered a similar situation in the case of nonlinear inversion by inverse Born series~\cite{markel_03_2}. As in the case of the small target, the reason for the incorrect reconstruction of the larger inhomogeneity is the incorrect assignment of noninteracting voxels. However, the reconstruction is not yet entirely broken for the contrast $\chi_0=1$. At $\chi_0=2$, too many noninteracting voxels have been assigned incorrectly and the resulting reconstruction is not useful.

\begin{figure}
% FIG 5
\centerline{\includegraphics[width=8.2cm]{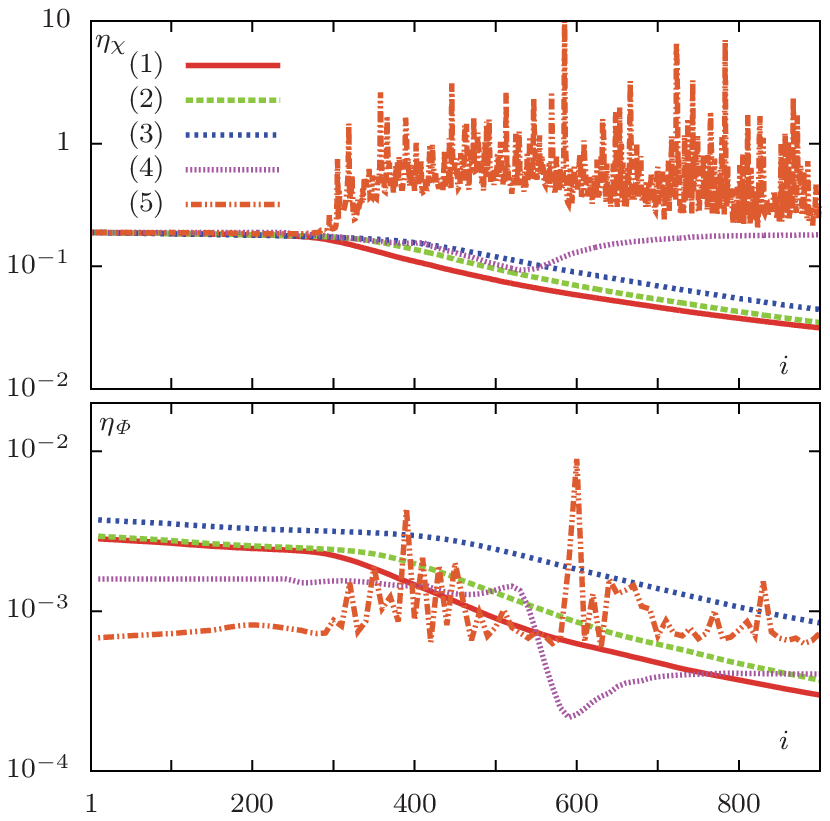}}
\caption{\label{fig:large_conv} Convergence data for the large target. The different line types correspond to different contrast $\chi_0$ as follows: $\chi_0=0.002$ (1), $\chi_0=0.02$ (2), $\chi_0=0.2$ (3), $\chi_0=1$ (4) and $\chi_0=2$ (5).}
\end{figure}

Convergence data for the large target are shown in Fig.~\ref{fig:large_conv}. The behavior of the errors is qualitatively similar to what was observed in the case of the small target, although the overall rate of convergence is obviously lower. We note that the results for $\chi_0=1$ and $\chi_0=2$ are qualitatively different from those for smaller values of the contrast. This is explained by the incorrect assignment of noninteracting voxels at the higher levels of contrast.

We note that the curve $\eta_{\it \Phi}(i)$ for the borderline case $\chi_0=1$ has a well pronounced minimum at $i=590$. We have observed a similar non-monotonous convergence for the smaller target as well. We have compared the reconstructions of the large target with $\chi_0=1$ obtained at the ``optimal'' iteration index $i=590$ and at $i=900$ 9data not shown). As expected, the result is visibly but not dramatically better at $i=590$. This result confirms our conjecture that monitoring the error $\eta_{\it \Phi}(i)$ is a useful approach to deciding when the iterations should stop.

\section{Methods to improve convergence}
\label{sec:impr}

This section investigates several improvements to the ``standard'' DCTMC algorithm. It will be shown that, with a few simple modifications, we can substantially reduce the number of iterations required to obtain accurate reconstructions. All simulations shown below concern the small target.

\subsection{Starting from the linear reconstruction}
\label{sec:impr.lin}

As was shown in~\cite{PRE_1}, DCTMC in the linear regime is equivalent to Richardson first-order iteration, which can have slow convergence. The regions of initial slow convergence are shown in Figs.~\ref{fig:small_conv} and \ref{fig:large_conv}. Two approaches can be used to cope with this problem. The first approach is to regularize the ISP as is explained in~\cite{PRE_1}. This, however, can result in image degradation beyond what is warranted by the intrinsic ill-posedness of the problem. The second approach is to start iterations with the linearized solution to the ISP, $V_{\rm lin}$, as the initial guess. This is computationally efficient because we have a fast linearized solver at our disposal as is explained in Sec.~\ref{sec:add.lin}.

\begin{figure}
% FIG 6
\centerline{\includegraphics[width=8.2cm]{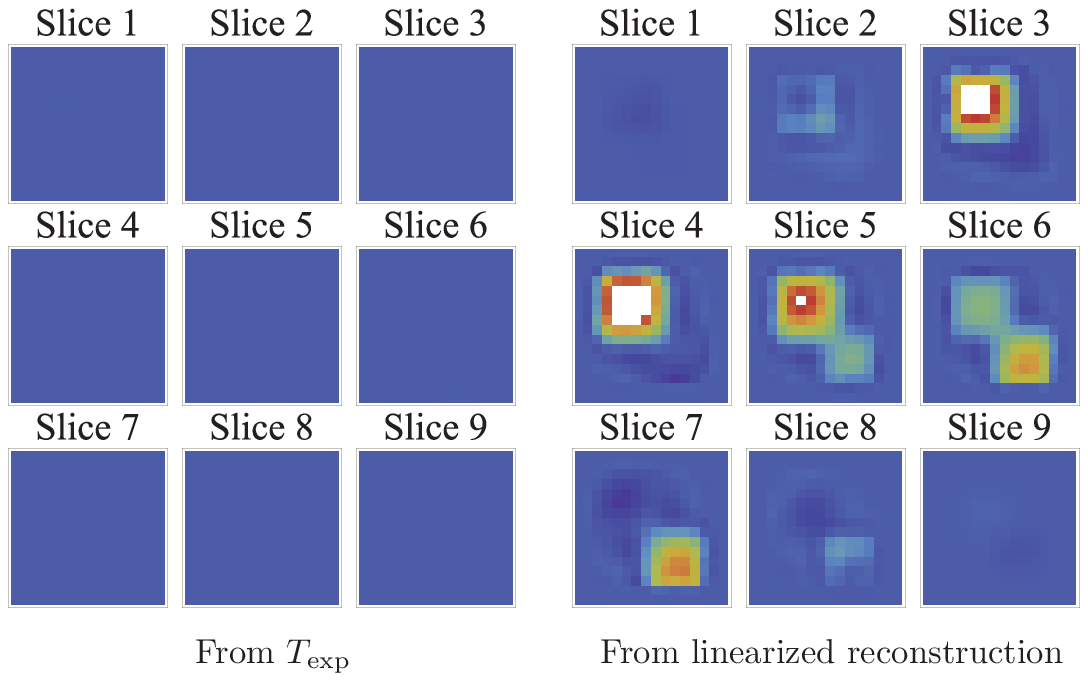}}
\caption{\label{fig:init_guess_comp} Two different initial guesses (DCTMC iteration starting points) for the small target with $\chi_0=0.175$. The left panel is the initial guess for the T-matrix, $T_1 = T_{\rm exp}$; the image is then computed by $V_1 = {\mathcal D}[(I + T_1{\it \Gamma})^{-1}T_1]$. The right panel is $V_1 = V_{\rm lin}$, where the linearized reconstruction $V_{\rm lin}$ was obtained by using the first Born approximation.}
\end{figure}

In Fig.~\ref{fig:init_guess_comp}, we compare two initial guesses. The first initial guess is obtained by $V_1 = {\mathcal D}[(I+T_{\rm exp}{\it \Gamma})^{-1}T_{\rm exp}]$. The second initial guess is $V_1 = V_{\rm lin}$. Both initial guesses were obtained for the small target and $\chi_0=0.175$. It can be seen that the first initial guess is very close to zero. It takes many Richardson iterations to transform it to the image resembling the one shown in the right panel. After this is accomplished, DCTMC starts to actually solve the nonlinear ISP. But the initial computational effort to arrive from the image in the left panel to the one in the right panel is unnecessary and can be avoided. 

To show that the number of necessary DCTMC iterations can be significantly reduced by using the initial guess $V_1 = V_{\rm lin}$, we have used the following algorithm:

\begin{enumerate}

\item[1:] Obtain the linearized reconstruction.

\item[2:] Using the result of Step 1 as the initial guess, run 5 iterations of DCTMC normally.

\item[3:] Then every 5 iterations check whether some of the values of $\chi_n$ satisfy $|\chi_n| < \chi_{\rm max}/500$, where $\chi_{\rm max} = \max_n|\chi_n|$.

\item[4:] If a given voxel satisfies the above condition 3 checks in a row, the corresponding $\chi_n$ is set to zero, and the computational domain is reduced.

\item[5:] The process is repeated with the following modifications. After $15$ iterations, checks are made every $20$ iterations and the relative threshold for determining a non-interacting voxel is reduced to the factor of $100$. After $200$ iterations, checks are made every $10$ iterations, and after $400$ iterations, the relative threshold is reduced to the factor of $60$, and after $600$ iterations the relative threshold is further reduced to the factor of $40$.

\end{enumerate}

Compared to the original algorithm described above, the interval between the sparsity checks is reduced at the beginning to utilize the non-interacting voxels found by the linear reconstruction, while simultaneously raising the relative threshold so as to not trust the linear reconstruction too much. Empirical evidence suggests that linearized reconstructions tend to capture the boundaries of an objects more or less correctly but the interior and the regions close to the boundaries can be reconstructed with a large error~\cite{markel_03_2}. Therefore, it can be expected that a linearized reconstruction would predict correctly the existing non-interacting voxels outside of the object and sufficiently far from its boundaries. However, noninteracting voxels in the interior or close to the boundaries can be predicted incorrectly. Therefore, we have modified the first $15$ iterations of the algorithm to avoid such incorrect assignments. After the initial $15$ iterations, the parameters of the algorithm are reset to match the original method. 

\begin{figure}
% FIG 7
\centerline{\includegraphics[width=8.2cm]{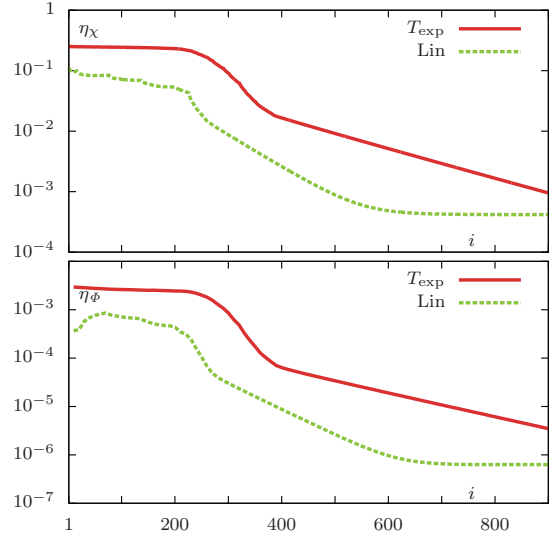}}
\caption{\label{fig:small_conv_linstart} Convergence rate for the small target with $\chi_0=0.175$. Different curves correspond to using the initial guess shown in Fig.~\ref{fig:init_guess_comp}, as labeled.}
\end{figure}

The error plots illustrating the convergence of DCTMC when started from the two different initial guesses are shown in Fig.~\ref{fig:small_conv_linstart}. A few immediate conclusions can be made from these plots. First, as we are starting from a much more reasonable guess, our initial error is much smaller. More importantly, as we can use sparsity checks accurately and relatively early in the iteration process, we have fast convergence almost from the start. Overall, for the contrast considered, we arrive at the final result in about $150$ iterations less. This is significant but not yet dramatic.

Also, the above improvement is not obtained for the strongest contrast we have considered, $\chi_0=1.75$. Here, when we start from the linearized reconstruction, we do not outperform the original method (data not shown). This can be understood qualitatively by noting that the linearized reconstruction at this contrast is very far from the true target. In fact, it is further away in the $L_2$ norm from the true target than the zero initial guess. Moreover, the linearized reconstruction is very strongly affected by the choice of the Tikhonov regularization parameter; an attempt to tune the latter to receive the best or most reasonable image quality can be counter-productive. More research is needed to understand how the best initial guess should be formed in the very strong nonlinearity regime. However, we will see that combining {\em all the improvements} discussed in this section in one algorithm makes DCTMC convergence fast and reliable even at this high value of the contrast.

\subsection{Using Reciprocity of Sources and Detectors}
\label{sec:impr.rec}

For a vast majority of physical equations of interest, all measurements are reciprocal. In particular, this is true for the scalar wave equation considered here. This means that interchanging the source and detector yields the same measurement. For this reason, making physical measurements with interchanged sources and detectors does not provide any additional information about the target. Mathematically, reciprocity of measurements is equivalent to symmetry of the T-matrix $T$. 

In spite of what has been said above, there exists a subtle point in DCTMC that makes explicitly accounting for the reciprocity useful. Namely, the experimental T-matrix $T_{\rm exp}$ is generally non-symmetric. Therefore, it does not account for the reciprocity of measurements. Consequently, including the data points with interchanged sources and detectors in the data matrix ${\it \Phi}$ will change $T_{\rm exp}$ and make it closer to being symmetric and closer to correspondence to a diagonal $V$. This account of reciprocity does not require any additional physical measurements. All that is needed is to include in ${\it \Phi}$ the data points that correspond to each source-detector pair being interchanged. In the traditional approaches to ISPs, such additional data points can be viewed as completely redundant and useless. However, in the context of DCTMC, inclusion of these data points makes $T_{\rm exp}$ more informative. In a sense, by including these redundant data points, we apply the additional condition that $T_{\rm exp}$ should be as symmetric as possible in the real space representation. 

\begin{figure}
% FIG 8
\centerline{\includegraphics[width=8.2cm]{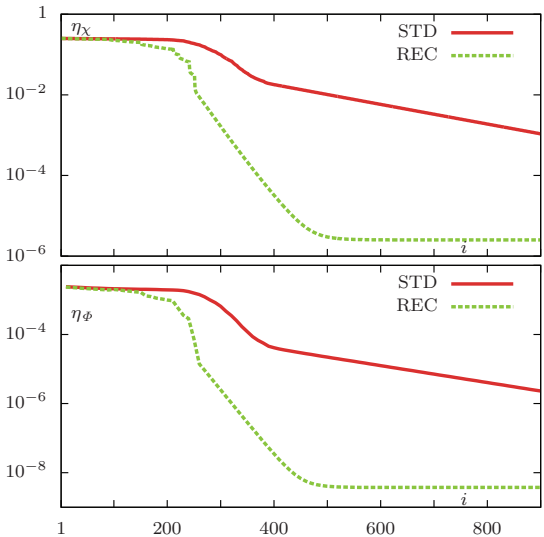}}
\caption{\label{fig:small_conv_rec} Convergence data for the small target with $\chi_0=0.0175$ comparing the original DCTMC algorithm and the one using reciprocity of sources and detectors.}
\end{figure}

We now investigate the doubling of the data set that is obtained by interchanging sources and detectors. In fact, the consequences of this operation can be quite dramatic. To illustrate the point, we show in Fig.~\ref{fig:small_conv_rec} the convergence plots for the small target at the contrast level $\chi_0=0.0175$. The initial guess $T_1 = T_{\rm exp}$ was used as the starting point. Two different data sets were used in the reconstructions. In the first case, sources are on one side of the sample and detectors are on the other side. In the second case, we supplement this data set by the ``reciprocal'' data points obtained by interchanging the source and the detector in each source-detector pair. The size of the data set is doubled by this procedure. We emphasize again that the addition of the ``reciprocal'' data points is not truly redundant in DCTMC. Indeed, the fraction of known (non-zero) entries in the experimental T-matrix $\tilde{T}_{\rm exp}$ (in the singular function representation) is $4\%$ and $17\%$ in the two cases, respectively. That is, by using reciprocity, the dimension of the non-zero minor in $\tilde{T}_{\rm exp}$ that is used in overwriting the iteratively updated T-matrix is roughly doubled. In other words, the experimental T-matrix is more informative in the second case.

It can be seen that the convergence speed is much higher with the use of reciprocity. Moreover, the converged results are much more precise with both errors $\eta_\chi$ and $\eta_{\it \Phi}$ approaching the limit determined by the numerical precision of the computer. This is a dramatic improvement for a simple change that does not require any additional information or processing power. The case represented in the Figure is for $\chi_0=0.0175$, but all other levels of contrast that we have considered exhibit a similar behavior.

\subsection{Using weighted summation to the diagonal for the operator ${\mathcal D}[\cdot]$}
\label{sec:impr.sum}

Next, we abandon the use of the Computational Shortcut 2 (defined in Ref.~\cite{PRE_1}) and use instead the force-diagonalization operator ${\mathcal D}[\cdot]$, viz, 
\begin{align}
\label{D_def}
\left({\cal D}[V] \right)_{ij} \equiv
\delta_{ij} \sum_k  V_{ik} h(\ell_{ki}) \ .
\end{align}
\noindent
Correspondingly, we will use the iteration scheme ``Main iteration without the use of Computational Shortcut 2,''~\cite{PRE_1}. This algorithm is more in line with the nonlocal framework of DCTMC. Although the computational time per one iteration will be increased by approximately the factor of $2$, we will see that the improvements obtained by this approach are the most promising in terms of significantly improving the convergence rate of DCTMC. It is worth emphasizing that the results in this sub-section are obtained without the improvements related to starting from a better initial guesses and from doubling the data set by interchanging sources and detectors. We will also not apply the sparsity checks. Therefore, the results shown below can be attributed to DCTMC in its most basic form. 

\begin{figure*}
% FIG 9
\centerline{\includegraphics[width=16.4cm]{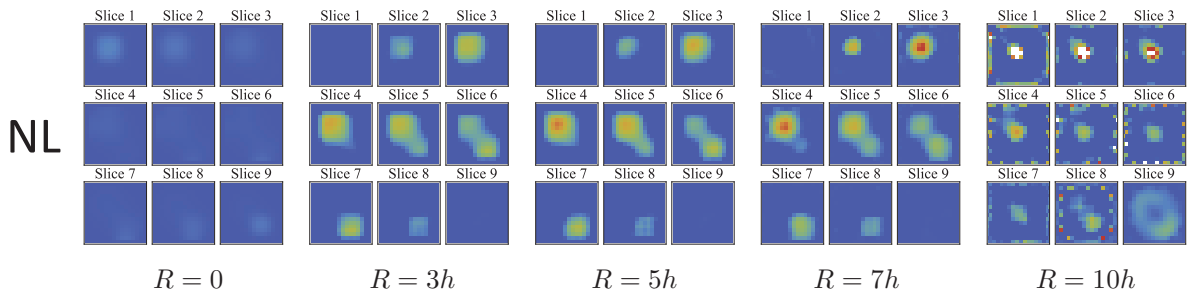}}
\caption{\label{fig:small_wsum} DCTMC reconstruction of the small target after $150$ iterations with varying degrees of $R$, where the row-summing is done over voxels separated by no more than $R$.}
\end{figure*}

As an example, we use a simple step-function for $\rho(\ell)$. That is, $\rho(\ell) = 1$ if $\ell \leq R$ and $\rho(\ell)=0$ otherwise. For the contrast $\chi_0=0.0175$, $150$ DCTMC iterations were performed without any sparsity checks. The reconstructions obtained after these $150$ iterations for varying values of $R$ are shown in Fig.~\ref{fig:small_wsum}. The immediate conclusion is that the case $R=5$ is much better than the original case $R=0$. It is also true that $R=10h$ is too large, coupling voxels that are too far apart, which creates a characteristic aliasing. 

\begin{figure}
% FIG 10
\centerline{\includegraphics[width=8.2cm]{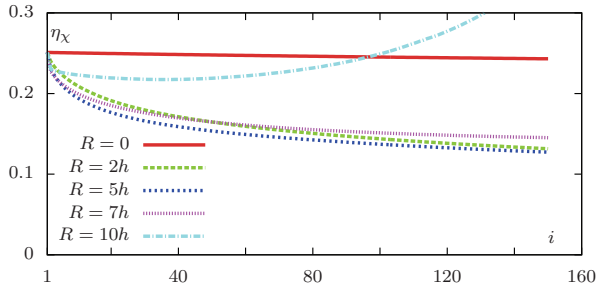}}
\caption{\label{fig:rowsum2} Convergence data $\eta_\chi$ for the reconstruction shown in Fig.~\ref{fig:small_wsum}.}
\end{figure}

The convergence data (for $\eta_\chi$ only and up to the iteration index $i=150$) are shown in Fig.~\ref{fig:rowsum2}. It can be seen that a dramatic improvement of the convergence speed is achieved by using weighted summation to the diagonal with the appropriate radius of influence. The case $R=5$ is optimal, with $R=2.5$ and $R=7.5$ being close in performance. Furthermore, it is interesting that for the case $R=10$, the first couple of iterations are quite effective, before something goes wrong and the error curve turns sharply upwards. This example is not suppressing as the nonlocal interaction of the far away voxels causes an unwanted behavior. Note that the case $R=0$ corresponds to the ``standard'' algorithm and the Computational Shortcut 2 was used for this case.  But despite the fact that the $R=0$ case completed in about half the time, looking at the error plot in Fig.~\ref{fig:rowsum2} shows that there is no confusion on the preference between $75$ iterations for $R=5$ and $150$ iterations of $R=0$.

\subsection{Putting it all together}

We now combining all three improvements described above in one algorithm. We will demonstrate that for all values of the contrast used for the small sample, we can achieve acceptable either perfect or at least acceptable results in $75$ iterations, compared to the $900$ iterations previously performed. The algorithm used in this section proceeds as follows:

\begin{enumerate}

\item[1:] Run the linear reconstruction.

\item[2:] Using the result of step 1 as the initial guess, run 5 iterations without sparsity checks.

\item[3:] Then every $5$ iterations check whether some susceptibilities $\chi_n$ satisfy $\vert \chi_n \vert < \chi_{\rm max}/500$, where $\chi_{\rm max} = \max_n \vert \chi_n \vert$.

\item[4:] If a given voxel satisfies the above condition 3 checks in a row, the corresponding $\chi_n$ is set to zero, and the computational domain is reduced.

\item[5:] The process is repeated with the following modifications. After $20$ iterations, the relative threshold for determining a non-interacting voxel is reduced to the factor of $100$. After $40$ iterations, the relative threshold is further reduced to the factor of $60$.

\end{enumerate}

\begin{figure*}
% FIG 11
\centerline{\includegraphics[width=16.4cm]{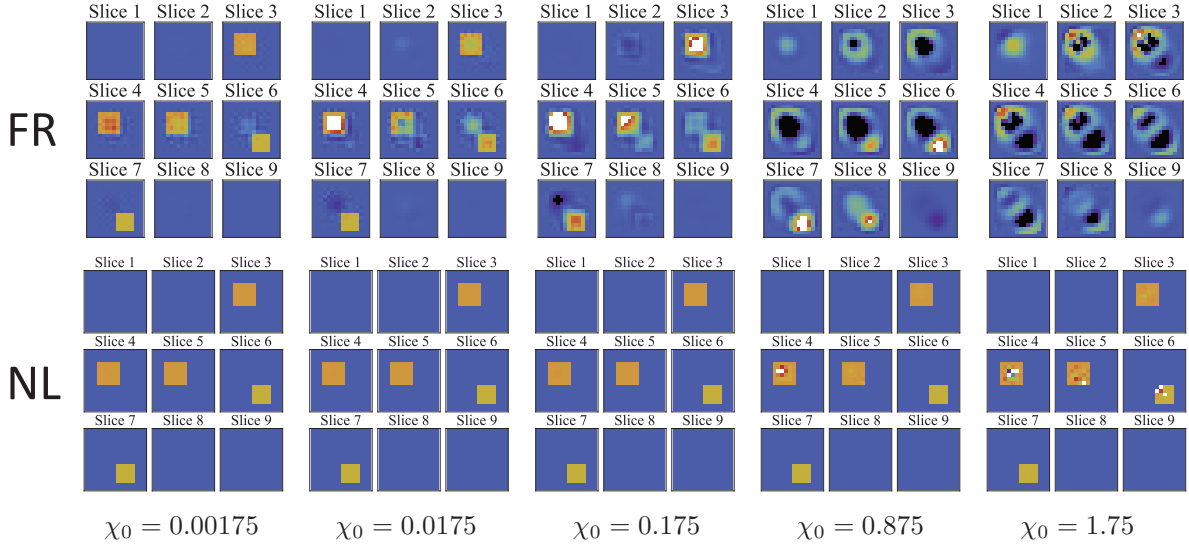}}
\caption{\label{fig:results} Reconstructions of the small target by the algorithm combining all three improvements discussed in this section after $75$ iterations for the five levels of contrast previously considered.}
\end{figure*}

Throughout the algorithm, reciprocity of sources and detectors was used, as well as weighted summation to the diagonal for the operator ${\mathcal D}[\cdot]$. For the weight function, we used the inverse distanced, namely, $\rho(\ell_{ik}) = \delta_{ik} + (1 - \delta_{ik})\ell_{ik}^{-1}$. Computational Shortcut 2 was not used. The reconstruction after $75$ iterations are shown in Fig.~\ref{fig:results} for all levels of contrast used previously. All reconstructions are nearly perfect at least for the largest value of the contrast, $\chi_0=1.75$. In the latter case, the reconstruction is not perfect but acceptable. The imprecision is again caused by three voxels that were incorrectly assigned as noninteracting. This happened even though we have increased the minimum relative threshold for determining noninteracting voxels from $40$ to $60$. 

\begin{figure}
% FIG 12
\centerline{\includegraphics[width=8.2cm]{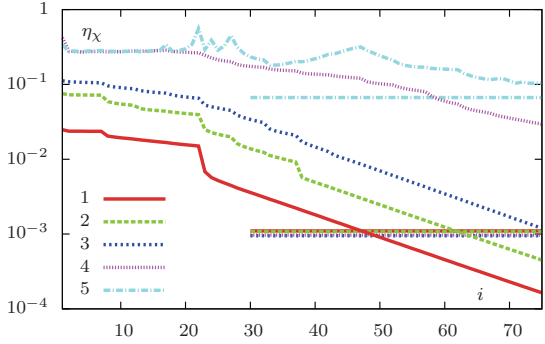}}
\caption{\label{fig:chi_plots} Convergence data of $\eta_\chi$ for the algorithm combining all improvements discussed in this section. The horizontal lines show the respective error attained after $900$ iterations of DCTMC without any of the improvements. Different line types correspond to different contrast $\chi_0$ as follows: $\chi_0=0.00175$ (1), $\chi_0=0.0175$ (2), $\chi_0=0.175$ (3), $\chi_0=0.875$ (4), and $\chi_0=1.75$ (5). }
\end{figure}

The convergence data for the algorithm employed in this subsection are shown in Fig.~\ref{fig:chi_plots}. It can be seen that the most dramatic improvement was obtained for the three lowest levels of contrast, as starting from the linearized inversion is of great assistance in these cases. In the intermediate case $\chi_0=0.875$, the reconstruction after $75$ iterations is not as precise as the one produced by the original method after $900$ iterations. This is clearly so because one voxel was erroneously assigned as noninteracting by the improved method. One incorrectly assigned noninteracting voxel does not degrade the reconstructed image seriously but bounds from below the error $\eta_\chi$ that can be reached. Clearly, more research on the application of sparsity checks in the strong nonlinearity regime is needed, especially when linearized inversion is used as the initial guess..

\section{Comparison with Gauss-Newton method}
\label{sec:GN}

For comparison purposes, we have also implemented the regularized Gauss-Newton method. We used the standard implementation of Levenberg-Marquardt with multiplicative damping. This method was too slow to be applied to the large target but we could apply it to the small target, even though one iteration of the Gauss-Newton method took much longer than one iteration of DCTMC. More importantly, the Gauss-Newton method did not converge to a reasonable result for the small target even at the lowest level of the contrast considered above. It is worth stressing that numerous modifications of the regularization and damping techniques were tried without producing any noticeably better results. Therefore, the Gauss-Newton reconstruction shown below is representative.

\begin{figure}
% FIG 13
\centerline{\includegraphics[width=4.1cm]{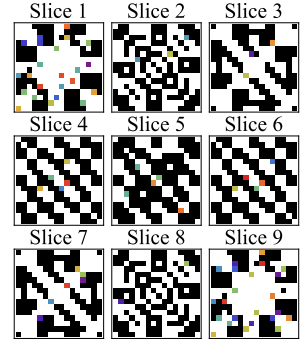}}
\caption{\label{fig:GN_1} A typical Gauss-Newton (with Levenburg-Marquardt damping technique) reconstruction of the small target with $\alpha_0=0.00175$.}
\end{figure}

\begin{figure}
% FIG 14
\centerline{\includegraphics[width=8.2cm]{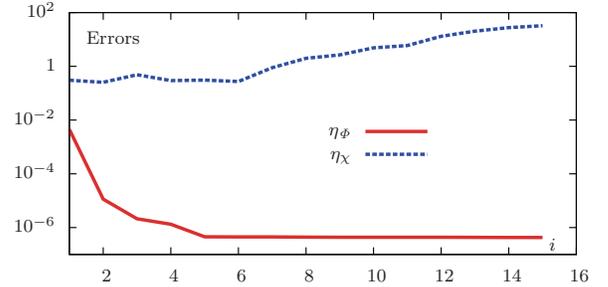}}
\caption{\label{fig:GN_1_plots} Convergence data for the Gauss-Newton (with Levenberg-Marquardt damping technique) reconstruction shown in Fig.~\ref{fig:GN_1}. Note that for $i>5$, the error $\eta_{\it \Phi}(i)$ continues to decrease, as is guaranteed by the algorithm, although this decrease is not visible due to the logarithmic vertical scale used in the figure.}
\end{figure}

A typical Gauss-Newton reconstruction of the small target with $\alpha_0=0.00175$ is shown in Fig.~\ref{fig:GN_1}. The corresponding error plots are shown in Fig.~\ref{fig:GN_1_plots}. It is obvious from the data of Fig.~\ref{fig:GN_1_plots} that the Gauss-Newton iterations are converging to a false solution (a local minimum of the cost function). Specifically, the error of the reconstruction $\eta_\chi$ is increasing with each iteration while the error of the equation $\eta_{\it \Phi}$ is decreasing and reaching some very small values that are consistent with the limit imposed by the finite numerical precision of the computer. 

\begin{figure}
% FIG 15
\centerline{\includegraphics[width=8.2cm]{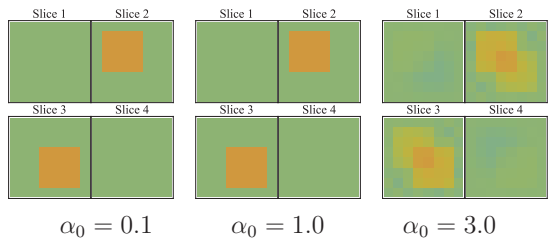}}
\caption{\label{fig:rec_0} Gauss-Newton (with Levenburg-Marquardt damping technique) reconstructions of the tiny target for different levels of contrast as labeled.}
\end{figure}

To confirm that the Gauss-Newton method was programmed correctly, we have reduced the size of the target even further to a point where the method works well, at least, for some levels of contrast. The shape of this even smaller (tiny) target is shown in Fig.~\ref{fig:Targets} and reconstructions are shown in Fig.~\ref{fig:rec_0}. For the tiny target, the Gauss-Newton reconstructions work well except at the high level of contrast ($\chi_0=3$), when the method fails to localize the inhomogeneity correctly.

Thus, we have produced numerical evidence that the DCTMC algorithm is able to solve nonlinear ISPs where other mainstream nonlinear inversion techniques fail. The results reported here further suggest that the size of the target is a critical factor. Specifically, the Gauss-Newton method works reasonably well for the tiny target but fails dramatically for the small target. A possible mathematical interpretation of this empirical observation is that the cost function used by the Gauss-Newton method develops a spurious local minimum for the small target. This minimum is connected by some path in the multi-dimensional space of solutions along which the cost function increases to the linearized solution. It follows therefore that the linearized solution is not a true minimum of the cost function. This statement may seem counterintuitive since we know for sure that the linearized solution minimizes the linearized cost function. However, the subtle but important point is that we do not really know whether the linearized solution minimizes the nonlinear cost function. It is also a realistic possibility that the linearized solution is in fact a minimum of the nonlinear cost function but this minimum is so shallow that, numerically, the iterations always overshoot and eventually descend towards the false solution. Clearly, more research in this direction is needed to fully understand the advantages and disadvantages of DCTMC compared to the standard methods.

\section{Discussion}
\label{sec:disc}

We have provided an initial numerical investigation of the data-compatible T-matrix completion (DCTMC) method for solving nonlinear inverse scattering problems (ISPs).
DCTMC was implemented for the scalar wave equation of the form \eqref{wave_1} which is relevant, in particular, to ultrasound imaging and seismic tomography. In many of the cases we have considered, DCTMC provided quantitatively accurate reconstructions whilst any linearized inversion (including those based on first Born, first Rytov or mean-field approximations) or the standard optimization methods (Gauss-Newton) have failed. This provides us with a proof-of-principle demonstration of the method's utility. Indeed, DCTMC does not utilize a cost function and does not suffer from the local minima and the associated false solution. Perhaps, this is the most important conceptual difference that distinguishes DCTMC from other nonlinear inversion methods. The method is especially useful for solving ISPs with large data sets. Even though we did not use supercomputers or any massively parallel computational platforms, we have succeeded in solving a strongly nonlinear inverse problem with more than $2 \cdot 10^6$ data points. Still, many questions regarding DCTMC improvement, optimization and applicability remain to be investigated either theoretically or numerically. Some of these topics for future research are listed below. 

First, the obvious task is to investigate DCTMC with adjustable sparsity checks. These checks are very useful for improving the speed and convergence of DCTMC. However, incorrect assignment of noninteracting voxels has resulted in the method breakup in the strong nonlinearity regime. We emphasize that all cases of DCTMC not providing a useful reconstruction were due to these incorrect noninteracting voxel assignments. If the sparsity checks are suppressed, the method can be slowly converging but we have not encountered and divergences or convergence to false solutions.

Second, we did not investigate in much detail various approaches to regularization. It is obvious that, as the target size increases, the ISP becomes more ill-posed. We can attempt a combination of Tikhonov regularization of the type that was discussed in~\cite{PRE_1}, physical constraints and, potentially, other methods. However, complicated interaction between different types of regularization require a careful and systematic investigation.

Third, there are questions of computational efficiency and inclusion of larger targets. A combination of methods described in Sec.~\ref{sec:impr} is expected to increase the rate of convergence or reduce the number of necessary iterations. Improving the computational bottlenecks is another important goal for future research. We note that some approximate approaches to speeding up this operations have been proposed by Jakobsen and Ursin~\cite{jakobsen_15_1} and these approaches can be utilized by DCTMC.

Fourth, weight functions $\rho(\ell)$ used in the definition of the force-diagonalization operator ${\mathcal D}[\cdot]$ require a separate investigation

We finally note that the major challenge in understanding and optimizing the DCTMC is the complicated interplay of the various tunable parameters and algorityhms thart are inherent to the method. Nevertheless, the authors are optimistic that further progress can be made by a combination of analytical and numerical investigations.

\section*{Acknowledgments}

This work has been carried out thanks to the support of the A*MIDEX project (No. ANR-11-IDEX-0001-02) funded by the ``Investissements d'Avenir'' French Government program, managed by the French National Research Agency (ANR) and was also supported by the US National Science Foundation under Grant DMS 1115616 and by the US National Institutes of health under Grant P41 RR002305. The authors are grateful to J.C.Schotland, A.Yodh and A.~Sentenac for very useful discussions.

\bibliographystyle{apsrev}
\bibliography{abbrev,book,master,local}

\begin{thebibliography}{25}
\expandafter\ifx\csname natexlab\endcsname\relax\def\natexlab#1{#1}\fi
\expandafter\ifx\csname bibnamefont\endcsname\relax
  \def\bibnamefont#1{#1}\fi
\expandafter\ifx\csname bibfnamefont\endcsname\relax
  \def\bibfnamefont#1{#1}\fi
\expandafter\ifx\csname citenamefont\endcsname\relax
  \def\citenamefont#1{#1}\fi
\expandafter\ifx\csname url\endcsname\relax
  \def\url#1{\texttt{#1}}\fi
\expandafter\ifx\csname urlprefix\endcsname\relax\def\urlprefix{URL }\fi
\providecommand{\bibinfo}[2]{#2}
\providecommand{\eprint}[2][]{\url{#2}}

\bibitem[{\citenamefont{Levinson and Markel}(2016)}]{PRE_1}
\bibinfo{author}{\bibfnamefont{H.~W.} \bibnamefont{Levinson}} \bibnamefont{and}
  \bibinfo{author}{\bibfnamefont{V.~A.} \bibnamefont{Markel}},
  \bibinfo{journal}{Phys. Rev. E}  (\bibinfo{year}{2016}),
  \bibinfo{note}{submitted}.

\bibitem[{\citenamefont{Bronstein et~al.}(2002)\citenamefont{Bronstein,
  Bronstein, Zibulevski, and Azhari}}]{bronstein_02_1}
\bibinfo{author}{\bibfnamefont{M.~M.} \bibnamefont{Bronstein}},
  \bibinfo{author}{\bibfnamefont{A.~M.} \bibnamefont{Bronstein}},
  \bibinfo{author}{\bibfnamefont{M.}~\bibnamefont{Zibulevski}},
  \bibnamefont{and} \bibinfo{author}{\bibfnamefont{H.}~\bibnamefont{Azhari}},
  \bibinfo{journal}{IEEE Trans. Med. Imag.} \textbf{\bibinfo{volume}{21}},
  \bibinfo{pages}{1395} (\bibinfo{year}{2002}).

\bibitem[{\citenamefont{DeVore and Polanko}(2005)}]{devore_05_1}
\bibinfo{author}{\bibfnamefont{G.~R.} \bibnamefont{DeVore}} \bibnamefont{and}
  \bibinfo{author}{\bibfnamefont{B.}~\bibnamefont{Polanko}},
  \bibinfo{journal}{J. Ultrasound} \textbf{\bibinfo{volume}{24}},
  \bibinfo{pages}{1685} (\bibinfo{year}{2005}).

\bibitem[{\citenamefont{Liu and Gu}(2012)}]{liu_12_1}
\bibinfo{author}{\bibfnamefont{Q.}~\bibnamefont{Liu}} \bibnamefont{and}
  \bibinfo{author}{\bibfnamefont{Y.~J.} \bibnamefont{Gu}},
  \bibinfo{journal}{Tectonophysics} \textbf{\bibinfo{volume}{16}},
  \bibinfo{pages}{31} (\bibinfo{year}{2012}).

\bibitem[{\citenamefont{Jakobsen}(2012)}]{jakobsen_12_1}
\bibinfo{author}{\bibfnamefont{M.}~\bibnamefont{Jakobsen}},
  \bibinfo{journal}{Stud. Geophys. Geod.} \textbf{\bibinfo{volume}{56}},
  \bibinfo{pages}{1} (\bibinfo{year}{2012}).

\bibitem[{\citenamefont{Jakobsen and Ursin}(2015)}]{jakobsen_15_1}
\bibinfo{author}{\bibfnamefont{M.}~\bibnamefont{Jakobsen}} \bibnamefont{and}
  \bibinfo{author}{\bibfnamefont{B.}~\bibnamefont{Ursin}}, \bibinfo{journal}{J.
  Geophys. Eng.} \textbf{\bibinfo{volume}{12}}, \bibinfo{pages}{400}
  (\bibinfo{year}{2015}).

\bibitem[{\citenamefont{Carney et~al.}(2004)\citenamefont{Carney, Frazin,
  Bozhevolnyi, Volkov, Boltasseva, and Schotland}}]{carney_04_1}
\bibinfo{author}{\bibfnamefont{P.~S.} \bibnamefont{Carney}},
  \bibinfo{author}{\bibfnamefont{R.~A.} \bibnamefont{Frazin}},
  \bibinfo{author}{\bibfnamefont{S.~I.} \bibnamefont{Bozhevolnyi}},
  \bibinfo{author}{\bibfnamefont{V.~S.} \bibnamefont{Volkov}},
  \bibinfo{author}{\bibfnamefont{A.}~\bibnamefont{Boltasseva}},
  \bibnamefont{and} \bibinfo{author}{\bibfnamefont{J.~C.}
  \bibnamefont{Schotland}}, \bibinfo{journal}{Phys. Rev. Lett.}
  \textbf{\bibinfo{volume}{92}}, \bibinfo{pages}{163903}
  (\bibinfo{year}{2004}).

\bibitem[{\citenamefont{Belkebir et~al.}(2005)\citenamefont{Belkebir, Chaumet,
  and Sentenac}}]{belkebir_05_1}
\bibinfo{author}{\bibfnamefont{K.}~\bibnamefont{Belkebir}},
  \bibinfo{author}{\bibfnamefont{P.~C.} \bibnamefont{Chaumet}},
  \bibnamefont{and} \bibinfo{author}{\bibfnamefont{A.}~\bibnamefont{Sentenac}},
  \bibinfo{journal}{J. Opt. Soc. Am. A} \textbf{\bibinfo{volume}{22}},
  \bibinfo{pages}{1889} (\bibinfo{year}{2005}).

\bibitem[{\citenamefont{Belkebir et~al.}(2006)\citenamefont{Belkebir, Chaumet,
  and Sentenac}}]{belkebir_06_1}
\bibinfo{author}{\bibfnamefont{K.}~\bibnamefont{Belkebir}},
  \bibinfo{author}{\bibfnamefont{P.~C.} \bibnamefont{Chaumet}},
  \bibnamefont{and} \bibinfo{author}{\bibfnamefont{A.}~\bibnamefont{Sentenac}},
  \bibinfo{journal}{J. Opt. Soc. Am. A} \textbf{\bibinfo{volume}{23}},
  \bibinfo{pages}{586} (\bibinfo{year}{2006}).

\bibitem[{\citenamefont{Bao and Li}(2007)}]{bao_07_1}
\bibinfo{author}{\bibfnamefont{G.}~\bibnamefont{Bao}} \bibnamefont{and}
  \bibinfo{author}{\bibfnamefont{P.}~\bibnamefont{Li}}, \bibinfo{journal}{Opt.
  Lett.} \textbf{\bibinfo{volume}{32}}, \bibinfo{pages}{1465}
  (\bibinfo{year}{2007}).

\bibitem[{\citenamefont{Mudry et~al.}(2012)\citenamefont{Mudry, Chaumet,
  Belkebir, and Sentenac}}]{mudry_12_1}
\bibinfo{author}{\bibfnamefont{E.}~\bibnamefont{Mudry}},
  \bibinfo{author}{\bibfnamefont{P.~C.} \bibnamefont{Chaumet}},
  \bibinfo{author}{\bibfnamefont{K.}~\bibnamefont{Belkebir}}, \bibnamefont{and}
  \bibinfo{author}{\bibfnamefont{A.}~\bibnamefont{Sentenac}},
  \bibinfo{journal}{Inverse Problems} \textbf{\bibinfo{volume}{28}},
  \bibinfo{pages}{065007} (\bibinfo{year}{2012}).

\bibitem[{\citenamefont{Boas et~al.}(2001)\citenamefont{Boas, Brooks, Miller,
  DiMarzio, Kilmer, Gaudette, and Zhang}}]{boas_01_1}
\bibinfo{author}{\bibfnamefont{D.~A.} \bibnamefont{Boas}},
  \bibinfo{author}{\bibfnamefont{D.~H.} \bibnamefont{Brooks}},
  \bibinfo{author}{\bibfnamefont{E.~L.} \bibnamefont{Miller}},
  \bibinfo{author}{\bibfnamefont{C.~A.} \bibnamefont{DiMarzio}},
  \bibinfo{author}{\bibfnamefont{M.}~\bibnamefont{Kilmer}},
  \bibinfo{author}{\bibfnamefont{R.~J.} \bibnamefont{Gaudette}},
  \bibnamefont{and} \bibinfo{author}{\bibfnamefont{Q.}~\bibnamefont{Zhang}},
  \bibinfo{journal}{IEEE Signal Proc. Mag.} \textbf{\bibinfo{volume}{18}},
  \bibinfo{pages}{57} (\bibinfo{year}{2001}).

\bibitem[{\citenamefont{Arridge and Schotland}(2009)}]{arridge_09_1}
\bibinfo{author}{\bibfnamefont{S.~R.} \bibnamefont{Arridge}} \bibnamefont{and}
  \bibinfo{author}{\bibfnamefont{J.~C.} \bibnamefont{Schotland}},
  \bibinfo{journal}{Inverse Problems} \textbf{\bibinfo{volume}{25}},
  \bibinfo{pages}{123010} (\bibinfo{year}{2009}).

\bibitem[{\citenamefont{Purcell and Pennypacker}(1973)}]{purcell_73_1}
\bibinfo{author}{\bibfnamefont{E.~M.} \bibnamefont{Purcell}} \bibnamefont{and}
  \bibinfo{author}{\bibfnamefont{C.~R.} \bibnamefont{Pennypacker}},
  \bibinfo{journal}{Astrophys. J.} \textbf{\bibinfo{volume}{186}},
  \bibinfo{pages}{705} (\bibinfo{year}{1973}).

\bibitem[{\citenamefont{Draine and Flatau}(1994)}]{draine_94_1}
\bibinfo{author}{\bibfnamefont{B.}~\bibnamefont{Draine}} \bibnamefont{and}
  \bibinfo{author}{\bibfnamefont{P.}~\bibnamefont{Flatau}},
  \bibinfo{journal}{J. Opt. Soc. Am. A} \textbf{\bibinfo{volume}{11}},
  \bibinfo{pages}{1491} (\bibinfo{year}{1994}).

\bibitem[{\citenamefont{Markel and Schotland}(2007)}]{markel_07_3}
\bibinfo{author}{\bibfnamefont{V.~A.} \bibnamefont{Markel}} \bibnamefont{and}
  \bibinfo{author}{\bibfnamefont{J.~C.} \bibnamefont{Schotland}},
  \bibinfo{journal}{Inverse Problems} \textbf{\bibinfo{volume}{23}},
  \bibinfo{pages}{1445} (\bibinfo{year}{2007}).

\bibitem[{fn1()}]{fn1}
\bibinfo{note}{One can envisage a situation in which the first equality in
  \eqref{disc-approx} is not an approximation but holds exactly. However, this
  can not be known {\em a priori} when solving an ISP. On the other hand, the
  second equality in \eqref{disc-approx} is never exact.}

\bibitem[{\citenamefont{Smunev et~al.}(2015)\citenamefont{Smunev, Chaumet, and
  Yurkin}}]{smunev_15_1}
\bibinfo{author}{\bibfnamefont{D.~A.} \bibnamefont{Smunev}},
  \bibinfo{author}{\bibfnamefont{P.~C.} \bibnamefont{Chaumet}},
  \bibnamefont{and} \bibinfo{author}{\bibfnamefont{M.~A.}
  \bibnamefont{Yurkin}}, \bibinfo{journal}{J. Quant. Spectrosc. Radiat.
  Transfer} \textbf{\bibinfo{volume}{156}}, \bibinfo{pages}{67}
  (\bibinfo{year}{2015}).

\bibitem[{\citenamefont{Draine}(1988)}]{draine_88_1}
\bibinfo{author}{\bibfnamefont{B.~T.} \bibnamefont{Draine}},
  \bibinfo{journal}{Astrophys. J.} \textbf{\bibinfo{volume}{333}},
  \bibinfo{pages}{848} (\bibinfo{year}{1988}).

\bibitem[{\citenamefont{Markel}(1992)}]{markel_92_1}
\bibinfo{author}{\bibfnamefont{V.~A.} \bibnamefont{Markel}},
  \bibinfo{journal}{J. Mod. Opt.} \textbf{\bibinfo{volume}{39}},
  \bibinfo{pages}{853} (\bibinfo{year}{1992}).

\bibitem[{\citenamefont{Lakhtakia}(1992)}]{lakhtakia_92_2}
\bibinfo{author}{\bibfnamefont{A.}~\bibnamefont{Lakhtakia}},
  \bibinfo{journal}{Optik} \textbf{\bibinfo{volume}{91}}, \bibinfo{pages}{134}
  (\bibinfo{year}{1992}).

\bibitem[{\citenamefont{Draine and Goodman}(1993)}]{draine_93_1}
\bibinfo{author}{\bibfnamefont{B.~T.} \bibnamefont{Draine}} \bibnamefont{and}
  \bibinfo{author}{\bibfnamefont{J.}~\bibnamefont{Goodman}},
  \bibinfo{journal}{Astrophys. J.} \textbf{\bibinfo{volume}{405}},
  \bibinfo{pages}{685} (\bibinfo{year}{1993}).

\bibitem[{\citenamefont{Yurkin}(2013)}]{yurkin_13_1}
\bibinfo{author}{\bibfnamefont{M.~A.} \bibnamefont{Yurkin}},
  \emph{\bibinfo{title}{{Handbook of Molecular Plasmonics}}}
  (\bibinfo{publisher}{Pan Stanford Pub.}, \bibinfo{address}{Singapore},
  \bibinfo{year}{2013}), chap. \bibinfo{chapter}{Computational approaches for
  plasmonics}, pp. \bibinfo{pages}{83--135}.

\bibitem[{\citenamefont{Ban et~al.}(2013)\citenamefont{Ban, Busch, Pathak,
  Moscatelli, Machida, C., Markel, and Yodh}}]{ban_13_1}
\bibinfo{author}{\bibfnamefont{H.~Y.} \bibnamefont{Ban}},
  \bibinfo{author}{\bibfnamefont{D.~R.} \bibnamefont{Busch}},
  \bibinfo{author}{\bibfnamefont{S.}~\bibnamefont{Pathak}},
  \bibinfo{author}{\bibfnamefont{F.~A.} \bibnamefont{Moscatelli}},
  \bibinfo{author}{\bibfnamefont{M.}~\bibnamefont{Machida}},
  \bibinfo{author}{\bibfnamefont{S.~J.} \bibnamefont{C.}},
  \bibinfo{author}{\bibfnamefont{V.~A.} \bibnamefont{Markel}},
  \bibnamefont{and} \bibinfo{author}{\bibfnamefont{A.~G.} \bibnamefont{Yodh}},
  \bibinfo{journal}{J. Biomed. Opt.} \textbf{\bibinfo{volume}{18}},
  \bibinfo{pages}{026016} (\bibinfo{year}{2013}).

\bibitem[{\citenamefont{Markel et~al.}(2003)\citenamefont{Markel, O'Sullivan,
  and Schotland}}]{markel_03_2}
\bibinfo{author}{\bibfnamefont{V.~A.} \bibnamefont{Markel}},
  \bibinfo{author}{\bibfnamefont{J.~A.} \bibnamefont{O'Sullivan}},
  \bibnamefont{and} \bibinfo{author}{\bibfnamefont{J.~C.}
  \bibnamefont{Schotland}}, \bibinfo{journal}{J. Opt. Soc. Am. A}
  \textbf{\bibinfo{volume}{20}}, \bibinfo{pages}{903} (\bibinfo{year}{2003}).

\end{thebibliography}

\end{document}